\documentclass[letterpaper,twocolumn,11pt,accepted=2022-07-03]{quantumarticle}

\pdfoutput=1

\usepackage{url}
\usepackage{ifthen}
\usepackage{cite}
\usepackage[cmex10]{amsmath} 
\usepackage{longtable}
\usepackage{makecell}
\usepackage{algorithm,algorithmic}
\usepackage{color}
\usepackage[normalem]{ulem}
\usepackage{amssymb,amsthm}
\usepackage{mathtools,bbm}  
                        
\usepackage{xifthen,xparse}
\usepackage{hyperref}  
\usepackage{float}
\usepackage{hhline}
\usepackage{caption}
\usepackage{subcaption}
\usepackage{physics}
\usepackage{stmaryrd}

\usepackage[table]{xcolor}

\usepackage{textcomp}

\allowdisplaybreaks

\usepackage[T1]{fontenc} 

\newtheorem{example}{Example}

\newenvironment{example*}
  {\addtocounter{example}{-1}\example}
  {\endexample}

\NewDocumentCommand\dketbra{+m+g}{%
  \IfNoValueTF{#2}
    {\left\lvert #1 \right\rangle \left\langle #1 \right\vert}
  {\left\lvert #1 \right\rangle \left\langle #2 \right\rvert}%
}
\NewDocumentCommand\dbraket{+m+g}{%
  \IfNoValueTF{#2}
    {\left\langle #1 \vert #1 \right\rangle}
  {\left\langle #1 \vert #2 \right\rangle}%
}

\global\long\def\Pr{\mathbb{P}}

\newcommand{\MCC}{\mathcal{C}}

\newcommand{\llbr}{[\![}
\newcommand{\rrbr}{]\!]}

\newif\ifnotes
\notestrue

\usepackage{soul}
\setstcolor{red} 
\setul{0pt}{0.7pt} 

\begin{document}
\title{Finite Rate QLDPC-GKP Coding Scheme that Surpasses the CSS Hamming Bound}

\author[1]{Nithin Raveendran}
\email{nithin@arizona.edu} 
\author[1]{Narayanan Rengaswamy}
\email{narayananr@arizona.edu} 

\author[2]{Filip Rozp\k{e}dek}
\email{frozpedek@uchicago.edu}

\author[3]{Ankur Raina}
\email{ankur@iiserb.ac.in}

\author[2]{Liang Jiang}
\email{liangjiang@uchicago.edu}

\author[1]{Bane Vasi{\'c}}
\email{vasic@ece.arizona.edu}%

\affil[1]{\normalsize Department of Electrical and Computer Engineering, University of Arizona, Tucson, Arizona 85721, USA}

\affil[2]{Pritzker School of Molecular Engineering, The University of Chicago, Chicago, IL 60637, USA}
\affil[3]{Department of Electrical Engineering and Computer Sciences, Indian Institute of Science Education and Research, Bhopal, Madhya Pradesh 462066, India}

\maketitle
\begin{abstract}
Quantum error correction has recently been shown to benefit greatly from specific physical encodings of the code qubits.
In particular, several researchers have considered the individual code qubits being encoded with the continuous variable Gottesman-Kitaev-Preskill (GKP) code, and then imposed an outer discrete-variable code such as the surface code on these GKP qubits.
Under such a concatenation scheme, the analog information from the inner GKP error correction improves the noise threshold of the outer code.
However, the surface code has vanishing rate and demands a lot of resources with growing distance.
In this work, we concatenate the GKP code with generic quantum low-density parity-check (QLDPC) codes and demonstrate a natural way to exploit the GKP analog information in iterative decoding algorithms. 
We first show the noise thresholds for two lifted product QLDPC code families, and then show the improvements of noise thresholds when the iterative decoder - a hardware-friendly min-sum algorithm (MSA) - utilizes the GKP analog information. 
We also show that, when the GKP analog information is combined with a sequential update schedule for MSA, the scheme surpasses the well-known CSS Hamming bound for these code families.
Furthermore, we observe that the GKP analog information 
helps the iterative decoder in escaping harmful trapping sets in the Tanner graph of the QLDPC code, thereby eliminating or significantly lowering the error floor of the logical error rate curves.
Finally, we discuss new fundamental and practical questions that arise from this work on channel capacity under GKP analog information, and on improving decoder design and analysis.
\end{abstract}

\section{Introduction}
\label{sec:intro}

Quantum low-density parity-check (QLDPC) codes form a promising family of quantum error-correcting codes (QECCs) since they involve stabilizer checks of bounded and low weight, have minimum distance scaling better than the square root of the code length, and can be decoded efficiently using iterative decoding methods \cite{mackay_quantum,ReviewQLDPC_Breuckmann,gottesman_fault_tolerant_ldpc,hastings2020fiber,BeyondSqrtMinD_highDimExpanders_Zemor2020, panteleev2020quantumLinearMinD,breuckmann2020balanced,panteleev2021quantumLinearMinDLocalTestable,QuantumTannerCodes,Decoding_QuantumExpanderCodes,HighDimQuantumHPC_Pryadko19}.
A specific construction of QLDPC codes is the hypergraph product construction~\cite{hypergraphProductCodeTillich}, which combines two arbitrary classical linear codes and produces a QECC.
If the component classical codes are LDPC codes, then the output QECC is a QLDPC code.
The simplest instance of this construction is the standard surface code, which is obtained as the hypergraph product of two binary repetition codes.
In just the last year, there have been several generalizations of this construction, which have led to the fiber bundle codes \cite{hastings2020fiber}, lifted product codes \cite{panteleev2020quantumLinearMinD}, and balanced product codes \cite{breuckmann2020balanced}, all of them improving the scaling of the minimum distance of QLDPC codes with respect to increasing code length.

While these codes are developed agnostic to the structure of the physical qubits in the encoding, it has been recently shown that using code-concatenation with the continuous variable (CV) Gottesman-Kitaev-Preskill (GKP) code~\cite{Gottesman-pra01,fukui2017analog,Fukui-prx18,Vuillot-pra19,Wang-ms19,Terhal-qst20,Noh-pra20,Hanggli-pra20,Noh-arxiv21,Noh-phd21,Grimsmo-prxq21} as the inner code can significantly boost the performance of the outer discrete-variable code. 
The single-mode GKP code encodes information in a $2$-dimensional subspace of the CV infinite-dimensional Hilbert space of a single harmonic oscillator.
This subspace is stabilized by two commuting CV displacement operators, which act as stabilizers of the GKP qubit.
For a square-lattice based GKP code, a measurement of these stabilizers enables correction of any random displacement error in phase space with magnitude at most $\sqrt{\pi}/2$ along each of the two conjugate quadratures. 
Furthermore, the GKP syndrome obtained through these measurements is a real-valued number which can be used as \emph{analog} information for an outer code that combines several GKP qubits.
Specifically, the syndrome provides a \emph{qubit-specific} reliability in terms of the probability that the qubit has undergone an effective GKP-logical error.
This is distinct from \emph{soft} information in classical and quantum error correction, which only uses the channel reliability information and all bits or qubits are assumed to undergo the same channel.
Such concatenation of ``physical'' GKP qubits with an outer stabilizer code has proven to be very beneficial in correcting errors in such systems~\cite{fukui2017analog}, and the idea has also been applied in quantum repeater protocols for quantum communications~\cite{rozpkedek2021quantum}.
In particular, recent works have characterized error thresholds for the case when the outer code is a surface code~\cite{Fukui-prx18,Vuillot-pra19,Wang-ms19,Terhal-qst20,Noh-pra20,Hanggli-pra20,Noh-arxiv21,Noh-phd21}.
Besides the noisy GKP data qubits in these works, i.e., data qubits that are finitely squeezed, the GKP ancillary qubits used for syndrome measurements (of both the inner and outer codes) are also noisy due to finite squeezing, and the aforementioned thresholds are with respect to the squeezing parameter (which is assumed to be the same for both data and ancilla).
The increase of threshold with the aid of analog information clearly shows that error ambiguities can be resolved effectively using the GKP analog information.

However, the family of surface codes encodes a \emph{fixed} number of logical qubits, e.g., $1$ or $2$, in a growing number of physical qubits, with the minimum distance scaling as the square root of the code size.
As a result, the surface-GKP concatenation schemes represent one extreme of the tradeoff between rate and reliability.
In applications such as quantum communications, it is also important to ensure high throughput, which means the coding schemes must have high rate with good enough reliability.
Even for fault-tolerant quantum computing, if one can devise high rate codes with good minimum distance, then this can greatly reduce the resource overhead~\cite{gottesman_fault_tolerant_ldpc}.
Therefore, our interest here is to investigate the benefits of using analog information when the outer code (family) is of high rate and has minimum distance scaling better than the square root of the code size.

We study the concatenation of the GKP code with \emph{lifted product (LP)} QLDPC codes~\cite{panteleev2020quantumLinearMinD} rather than just the surface code.
These codes have better rates since they encode much more than one logical qubit, and they are equipped with fast iterative decoding algorithms such as belief propagation (BP).
In this first work concatenating GKP inner codes with outer Calderbank-Shor-Steane (CSS)-based~\cite{calderbank1996quantum_exists,Steane-physreva96} \emph{general} QLDPC codes, we investigate a simple noise model where all GKP state preparations are perfect, GKP ancillas are noiseless (infinitely squeezed), the physical GKP qubits of the outer code undergo a Gaussian random displacement in phase space, and syndromes are measured for the inner codes using Steane's method \cite{Steane-physreva96}. Once GKP syndromes are obtained, an appropriate correction is applied on each GKP qubit.
Since all ancillas used in syndrome measurements are perfect, after the GKP correction operation each GKP qubit has either $0$ effective displacement or a logical GKP error in phase space.
This provides the ``true'' error pattern for the outer code.
Since the Gaussian random displacement channel generates errors that are uncorrelated in the $Q$- and $P$-quadratures, and since we are considering the square-lattice GKP encoding, the effective GKP logical errors translate to uncorrelated ``physical'' $X$- and $Z$-errors for the outer QLDPC code \cite{Gottesman-pra01}.
The GKP analog information in our setting is the probability that each qubit has undergone an effective logical error after GKP error correction (Steane's syndrome extraction followed by feedback displacement on the data qubits).
As the outer code is CSS, we measure $X$- and $Z$-stabilizers separately, also using noiseless GKP ancillas, and combine this syndrome with the analog information to initialize the BP-based iterative decoder \cite{mackay_quantum}.
The iterative decoder is executed separately for $X$- and $Z$-errors, and the estimated error pattern is compared with the ``true'' error on the GKP qubits to determine if the outer decoding was correct.
Since conventional BP decoder is complex in its hardware implementation, we use its low-complexity version, called the \emph{min-sum algorithm (MSA)} (and its variants), which is widely deployed in classical error correction applications \cite{05CDEFH,HocevarLayered}.
Thus, our results provide insight on the performance of the outer code decoder that is also hardware-friendly.

Given the setting of our work, we summarize our main contributions and results here. 

\begin{enumerate}

\item We investigate a general concatenated-GKP coded scheme where the outer codes are LP-QLDPC codes~\cite{panteleev2020quantumLinearMinD} from two code families, both consisting of codes lifted from $(m_b,n_b)$-regular quasi-cyclic LDPC ``base'' codes. 
Here, $m_b$ and $n_b$ refer respectively to the variable node and check node degree in the factor graph of the base codes.
The rate of the lifted product codes is $r \geq \frac{(n_b - m_b)^2}{(n_b^2 + m_b^2)}$.
One family, named LP04, has $(m_b,n_b) = (3,4)$ and asymptotic rate $1/25 = 0.04$, while the other family, named LP118, has $(m_b,n_b) = (3,5)$ and asymptotic rate $4/34 \approx 0.118$.

\item We consider iterative decoding of these LP-QLDPC codes and make explicit use of the GKP analog information in initializing the MSA decoder. 
For each GKP qubit, we convert the probability of GKP-logical $X$ or GKP-logical $Z$ error (see Eq.~\eqref{eq:analogInfoprob}) into a log-likelihood ratio (LLR), and use this as the ``channel input'' at the corresponding variable node for the MSA decoder. 
We also compare the performance to the setting where we ignore the analog information.

\item We demonstrate the improvement in performance with analog information by calculating the noise threshold in terms of the standard deviation $\sigma$ of the Gaussian random displacement channel, with and without analog information. 
For the LP04 family, the threshold improves from $\sigma(\text{LP04}) = 0.505$ to $\sigma(\text{LP04a}) = 0.557$, and for the LP118 family, the threshold improves from $\sigma(\text{LP118}) = 0.495$ to $\sigma(\text{LP118a}) = 0.547$, where the additional `a' in the family names refers to the use of GKP analog information in the outer MSA decoder.

\item For non-degenerate CSS codes, which are constructed from two compatible classical codes, the Hamming sphere packing bound for classical codes implies an upper bound on their rates~\cite{Steane-rspa96,calderbank1996quantum_exists,Dennis-jmp02}.
This bound, which we refer to as the ``CSS Hamming bound'', is given by $C(p) = 1 - 2 h_2(p)$, where $p$ is the probability that a qubit undergoes an error 
\footnote{The actual bound is given in terms of the minimum distance $d$ of the code of length $n$, which we have reinterpreted here using the error rate, $p \approx d/2n$.}
and $h_2(p) \coloneqq - p \log_2(p) - (1-p) \log_2(1-p)$ is the binary entropy function.
We show that, in the concatenated QLDPC-GKP setting, the outer CSS-QLDPC codes are able to surpass this bound when the outer decoder (a) uses the GKP analog information and (b) executes a \emph{sequential (or layered)} node update schedule rather than the more common parallel (or flooding) schedule.
More precisely, we set $C(p) = 0.04$ or $0.118$, and show that the codes can correct errors beyond the limit $p$ implied by the CSS Hamming bound, where for the GKP code {with the noise coming from the Gaussian random displacement channel}, $p$ is related to $\sigma$ as 
\begin{equation}
\begin{aligned}
\label{eq:p_sigma_relation}
    p(\sigma) & = \sum_{l \in \mathbb{Z}} \int_{(4l+1)\sqrt{\pi}/2}^{(4l+3)\sqrt{\pi}/2} \frac{1}{\sqrt{2 \sigma^2}} e^{-\frac{x^2}{2\sigma^2}}\,dx \\& \leq   2 \int_{\sqrt{\pi}/2}^{\infty} \frac{1}{\sqrt{2 \sigma^2}} e^{-\frac{x^2}{2\sigma^2}}\,dx .
\end{aligned}
\end{equation}
For the purposes of the calculations in this paper, we find that the difference between the exact value and the upper bound stated above is of the order of $10^{-6}$ and hence negligible for our purposes.
Since the CSS Hamming bound applies to non-degenerate codes, the contribution of the degeneracy of the LP-QLDPC codes to this advantage remains to be understood further.

\item We assess the performance of our schemes by plotting the logical error rate against the standard deviation $\sigma$.
This is similar to the strategy used in the investigations of the surface-GKP concatenated schemes, except that there is only one logical qubit in those schemes.
Since LP04 and LP118 encode many logical qubits, a fair comparison to the surface-GKP schemes would require plotting the logical \emph{qubit} error rate, which is the average rate of (logical) errors \emph{per logical qubit}.
This means that whenever a miscorrection occurs after decoding, we need to determine which logical qubits are in error.
However, it turns out that our logical error rate and logical qubit error rate plots will not differ by much, due to a unique phenomenon in the MSA decoder.
The decoder has three possible output scenarios: it matches the syndrome and succeeds in correcting the error either exactly or with another error that differs from the true error only by a stabilizer, i.e., a degenerate error (successful), or it matches the syndrome but introduces a logical error (unsuccessful; miscorrection), or it is unable to find an error pattern matching the syndrome, in which case we declare {that} all qubits are in error (unsuccessful; failure).
In our experiments, most of the unsuccessful decoding attempts were due to failures and not miscorrections. 
Thus, whenever the syndrome is matched, the error is almost surely corrected properly, and miscorrections rarely occur.
In other words, it is highly likely that either all logical qubits are in error or none of them are in error after the MSA decoding procedure.

\item In standard LDPC and QLDPC decoding, the iterative decoder suffers from an ``error floor'' phenomenon, where in the regime of low noise the logical error rate does not vanish but saturates at a finite value.
This is because of small graph configurations, called \emph{trapping sets}, which include short cycles, that cause the decoder to fail at low error rates.
While this is typically overcome by carefully designing the code such that small trapping sets are avoided, here we show that the GKP analog information itself seems to eliminate the error floor problem.
We demonstrate this using a particular code from the LP04 family, but we emphasize that this is a preliminary observation that requires further investigation.
Intuitively, it appears that the ``analog'' knowledge about the error on each qubit seems to suffice to bias the decoder onto a particular pattern inside the TS.
\end{enumerate}

The organization of the paper is as follows. 
In Section \ref{Sec:GKPConcatenation}, we discuss the inner GKP code and the general concatenation scheme. 
Then, we discuss the construction of QLDPC codes and their decoding using syndrome-based iterative message passing decoders in Section \ref{Sec:QLDPC_Code}. 
Subsequently, in Section \ref{sec:SimResults}, we discuss the specific LP code construction used in this work and present the simulation results demonstrating the advantages of GKP analog information in the QLDPC-GKP concatenated schemes.
Finally, we provide concluding remarks and discuss future research directions in Section \ref{sec:Conclusion}, where we mention some fundamental questions that arise from this concatenation scheme.

\section{General Concatenated-GKP Scheme}
\label{Sec:GKPConcatenation}

\subsection{Gottesman-Kitaev-Preskill (GKP) Code}
The GKP code is a bosonic stabilizer code that encodes quantum information in the CV phase space of harmonic oscillators~\cite{Gottesman-pra01,Grimsmo-prxq21}. 
In this work, we consider single-mode square-lattice-based GKP code which is defined by the two commuting stabilizers:
\begin{equation}
\hat{S}_Q = \exp(i2\sqrt{\pi}\hat{Q}), \quad \hat{S}_P = \exp(-i2\sqrt{\pi}\hat{P}).
\label{eq:GKPstabilizers}
\end{equation}
Here $\hat{Q}$ and $\hat{P}$ correspond to the two conjugate quadrature operators defining the $(Q,P)$ - phase space of a single bosonic mode and they satisfy the canonical commutation relation $[\hat{Q}, \hat{P}] = i$. The two stabilizers $\hat{S}_Q$ and $\hat{S}_P$ act as displacement operators by $2\sqrt{\pi}$ along the $\hat{P}$ and $\hat{Q}$ quadratures, respectively. As they are periodic functions of the two quadratures, a measurement of each of the two stabilizers can also be seen as the measurement of the corresponding quadrature modulo $\sqrt{\pi}$ such that the measurement outcome will lie in the interval $[-\sqrt{\pi}/2, \sqrt{\pi}/2)$. Hence, the GKP code naturally arises from the observation that while the two canonical conjugate quadratures cannot be measured simultaneously, such a measurement becomes possible when the two quadratures are measured modulo a fixed interval.

\begin{figure}
\includegraphics[width = \columnwidth]{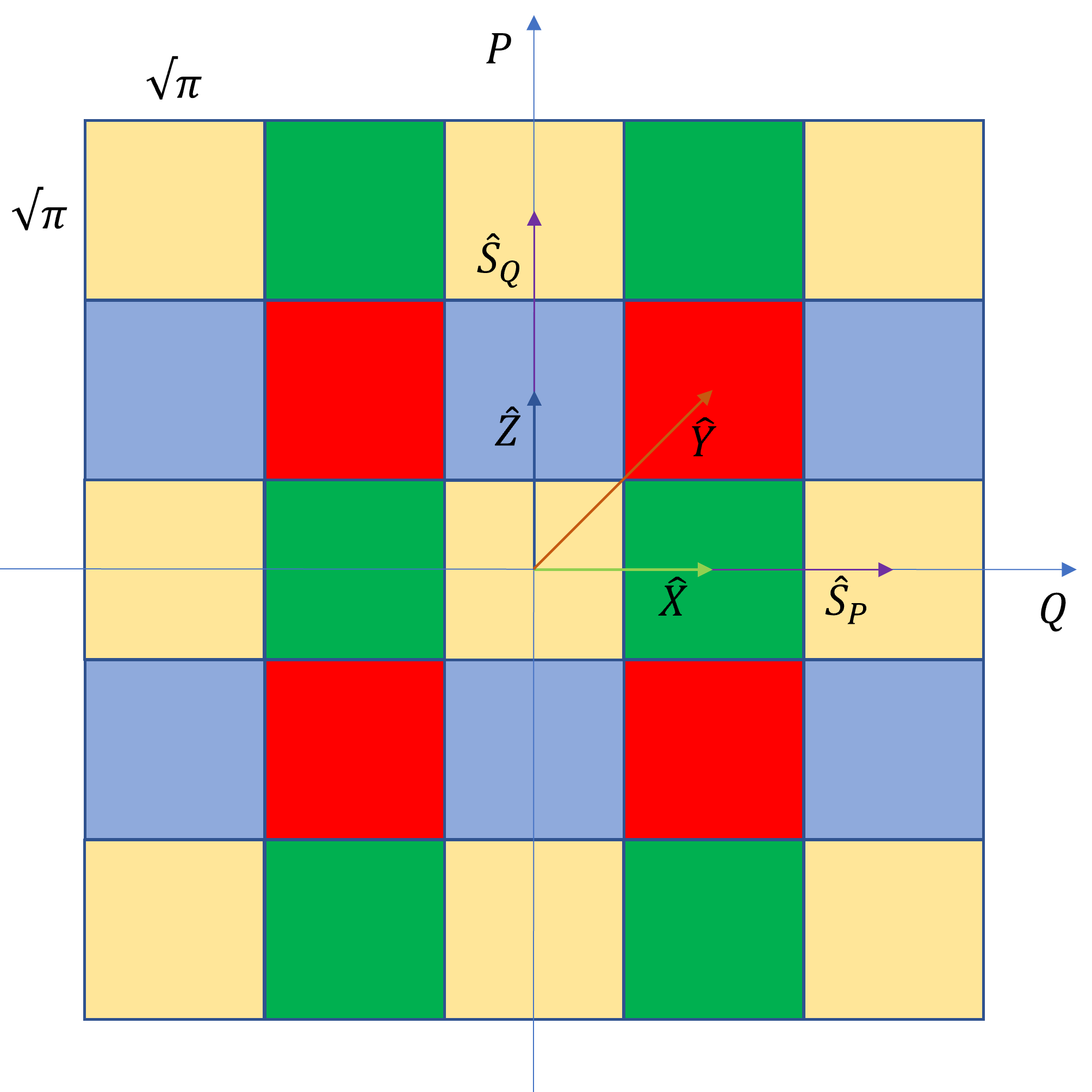}
\caption{Effect of different displacements in the $Q$-$P$ phase space on a logical qubit encoded in a square-lattice GKP code. Each of the marked square regions has a side length of $\sqrt{\pi}$. The action of the GKP stabilizers $\hat{S}_P$ and $\hat{S}_Q$, as well as the logical operators $\hat{X}/\hat{Z}/\hat{Y}$ is also marked. Displacement by $2\sqrt{\pi}$ along the $\hat{Q}$ or $\hat{P}$ quadrature corresponds to the GKP stabilizer and hence displacement to any of the yellow regions will result in successful error correction. Displacement to the green/blue/red region will result in a logical $X/Z/Y$ error respectively after error correction.}
\label{fig:PhaseSpace}
\end{figure}

For the square-lattice based GKP code, the logical $Z$ and $X$ operators are given by
\begin{align}
\begin{split}
& Z = \sqrt{\hat{S}_Q} = \exp(i\sqrt{\pi}\hat{Q}), \qquad \\& X = \sqrt{\hat{S}_P} = \exp(-i\sqrt{\pi}\hat{P}),
\end{split}
\end{align}
and the $Z$- and $X$-basis states can be expressed as
\begin{equation}
\begin{aligned}
&\ket{0_{\text{GKP}}} = \sum_{l \in \mathbb{Z}} \ket{Q = 2l\sqrt{\pi}}, \\&
\ket{1_{\text{GKP}}} = \sum_{l \in \mathbb{Z}} \ket{Q = (2l+1)\sqrt{\pi}}, \\&
\ket{+_{\text{GKP}}} = \sum_{l \in \mathbb{Z}} \ket{P = 2l\sqrt{\pi}}, \\&
\ket{-_{\text{GKP}}} = \sum_{l \in \mathbb{Z}} \ket{P = (2l+1)\sqrt{\pi}}.
\end{aligned}
\label{eq:GKPbasisstates}
\end{equation}
We see that while the two stabilizers enforce that the GKP states need to be invariant under displacement by $2\sqrt{\pi}$ in the two quadratures, the displacements by $\sqrt{\pi}$ implement logical $Z$ and $X$ operations. Moreover, since the GKP syndrome for each quadrature is a real number belonging to the interval $[-\sqrt{\pi}/2, \sqrt{\pi}/2)$, random shifts along the two axes that are larger in magnitude than $\sqrt{\pi}/2$ will result in most cases in logical errors after correction. Specifically, a displacement whose component along the $\hat{Q}$ ($\hat{P}$) quadrature lies in the interval $[(l-1/2) \sqrt{\pi},(l+1/2) \sqrt{\pi})$) leads to a logical $X$ ($Z$) error for odd values of $l$ and to no-error for even values of $l$. Since in most physically motivated scenarios the random displacement is governed by a Gaussian distribution with mean zero, the regime in which the GKP code allows for efficient suppression of errors corresponds to error channels for which a displacement belonging to the interval with $\abs{l}>0$ occurs with small probability. These observations lead precisely to the Eq.~\eqref{eq:p_sigma_relation} stated before. The effect of different displacements in phase space on the GKP-encoded qubit is shown in Fig.~\ref{fig:PhaseSpace}.

\begin{figure*}
\centering
\includegraphics[width = \textwidth]{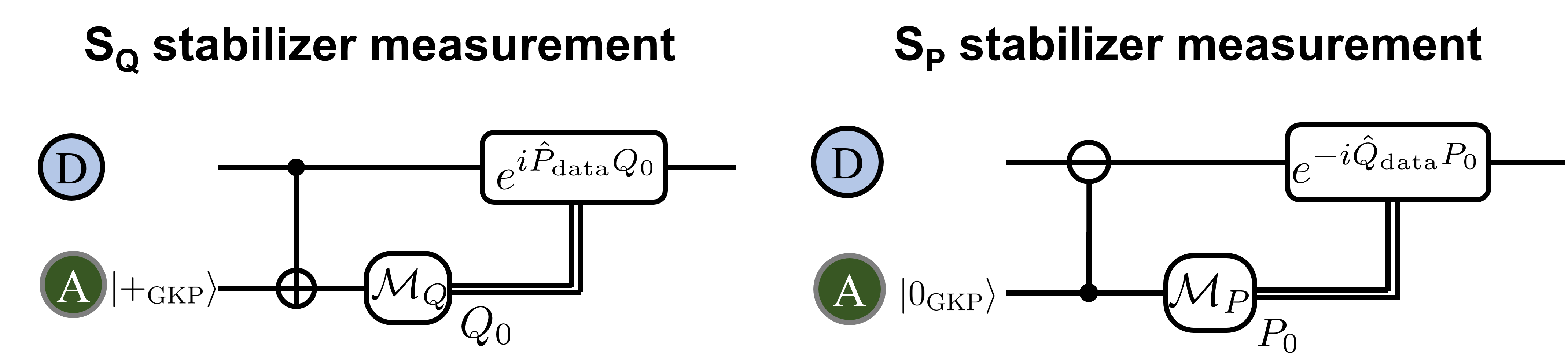}
\caption{GKP Steane error correction. Measurement of a single GKP stabilizer requires a single GKP ancilla. When measuring the $\hat{S}_{\text{Q}}$ ($\hat{S}_{\text{P}}$) stabilizer, the ancilla is initialised in the GKP logical plus (zero) state. Then the $\text{SUM}_{\text{D} \rightarrow \text{A}}$ gate defined in Eq.~\eqref{eq:SumGate} and marked as controlled-$\oplus$ is applied from the data qubit (marked as D) onto the ancilla (marked as A). The inverse SUM gate, $\text{SUM}_{\text{A} \rightarrow \text{D}}^\dag$, marked as controlled-$\ominus$ is applied from the ancilla onto the data qubit. After that the ancilla is measured in the $\hat{Q}$ ($\hat{P}$) quadrature. The outcome modulo $\sqrt{\pi}$ belonging to the interval $[-\sqrt{\pi}/2, \sqrt{\pi}/2)$ is denoted by $Q_0$ ($P_0$). Finally the correction operation is performed by applying the displacement operator $e^{i\hat{P}_{\text{data}}Q_0}$ ($e^{-i\hat{Q}_{\text{data}}P_0}$) which implements the transformation $\hat{Q}_{\text{data}} \rightarrow \hat{Q}_{\text{data}} - Q_0$ ($\hat{P}_{\text{data}} \rightarrow \hat{P}_{\text{data}} -P_0$). Figure taken with modifications from~\cite{rozpkedek2021quantum} under the license CC BY 4.0 https://creativecommons.org/licenses/by/4.0/.}
\label{fig:SyndromeInner}
\end{figure*}

Measuring the stabilizers given in Eq.~\eqref{eq:GKPstabilizers} and performing error correction requires additional ancillary GKP qubits. 
Two possible ways of implementing this procedure are Steane error correction~\cite{Gottesman-pra01} and teleportation-based error correction~\cite{walshe2020continuous,fukui2021all}. 
In the first method, depicted in Fig.~\ref{fig:SyndromeInner}, the measurements of the $\hat{S}_Q$ and $\hat{S}_P$ stabilizers are done separately and independently using one GKP ancilla for each measurement. After obtaining the stabilizer outcome from the measurement on the ancilla, a feedback displacement is applied to the data qubit to correct the errors in the given quadrature. 
In the teleportation-based error correction, a local GKP Bell pair between two GKP ancillas needs to be generated in advance. Then a quantum teleportation through such a GKP Bell pair, in which a Bell measurement on the GKP data qubit and one-half of the GKP Bell pair is performed, enables for GKP error correction. We note that in contrast to Steane error correction, in this case two GKP qubits are measured at once and therefore both stabilizers $\hat{S}_Q$ and $\hat{S}_P$ are measured simultaneously.
The performance of these two error correction schemes depends on the noise model, which for our setting is described in Section \ref{Sec:noiseModel}.

\subsection{Noise Model}
\label{Sec:noiseModel}

We assume that the individual GKP (data) qubits are subjected to a Gaussian random displacement channel~\cite{Gottesman-pra01}:
\begin{equation}
\mathcal{E}_{\text{disp}}[\sigma](\rho) = \frac{1}{\pi \sigma^2}\int d^2\alpha \exp\left[-\frac{\abs{\alpha}^2}{\sigma^2}\right] \hat{D}(\alpha)\rho \hat{D^\dag}(\alpha),
\label{eq:randomgaussiandisplChannel}
\end{equation}
where $\hat{D}(\alpha) = \exp[\alpha \hat{a}^\dag - \alpha^\ast \hat{a}]$ is the displacement operator, $\hat{a}^\dag$ is the conjugate tranpose of $\hat{a}$, $\alpha = \alpha_{\text{Re}} +  i \alpha_{\text{Im}}$, and $\alpha^\ast = \alpha_{\text{Re}} -  i \alpha_{\text{Im}}$. 
The bosonic annihilation and creation operators $\hat{a}$ and $\hat{a}^\dag$ satisfy the commutation relation $[\hat{a}^\dag, \hat{a}] = 1$ and the position and momentum operators can be expressed as: $\hat{Q} = (\hat{a}^\dag + \hat{a})/\sqrt{2}$ and $\hat{P} = i(\hat{a}^\dag - \hat{a})/\sqrt{2}$. 
Hence, the displacement in the $(Q,P)$ phase space by $(Q_0, P_0)$ corresponds to $Q_0 = \sqrt{2}\alpha_{\text{Re}}$ and $P_0 = \sqrt{2}\alpha_{\text{Im}}$.
Finally, $\sigma^2$ is the variance of the Gaussian noise distribution. 
This noise model is physically motivated. 
Specifically, the two most common bosonic noise channels are precisely the Gaussian random displacement channel and the photon loss channel. The latter one can in fact be converted into the former since supplementing the photon loss process with an additional step of phase-insensitive amplification, where gain is given by the inverse of the transmissivity of the loss channel, results exactly in a Gaussian random displacement channel~\cite{noh2018quantum, albert2018performance, kim1996quantum,sabapathy2011robustness,ivan2011operator}. 
Being able to mitigate the effect of photon loss is of great importance as this type of noise arises naturally in the communication setting of optical channels used for transmission of quantum information.
Moreover, microwave cavities, which provide for a powerful platform for processing quantum information, also suffer from the loss of microwave photons~\cite{ma2021quantum}.
Hence being able to overcome the errors introduced by the Gaussian random displacement channel is of great relevance both to quantum communication and to quantum computing.
This shows that using Gaussian random displacement as a model for our noise channel has a practical motivation --- it describes well the various physical processes to which GKP qubits could be subjected in different practical realizations.
In fact the GKP encoding and GKP error correction have already been experimentally realised in the platform of superconducting microwave cavity~\cite{campagne2020quantum} as well as in trapped ion mechanical oscillators~\cite{fluhmann2019encoding, de2020error}.

In practice, the ideal GKP states that are exactly stabilized by the operators in Eq.~\eqref{eq:GKPstabilizers}, and with basis states given in Eq.~\eqref{eq:GKPbasisstates}, are unphysical as they correspond to infinite amount of squeezing and hence require infinite amount of energy. 
Practical GKP states will have finite amount of squeezing and will be stabilized by the operators in Eq.~\eqref{eq:GKPstabilizers} only approximately. 
Furthermore, implementing operations between the GKP data and ancilla qubits as well as homodyne measurements performed on the GKP qubits can introduce additional errors.
Finally, the processes of photon loss and heating will in general be continuous processes that affect the idle GKP qubits also during the error correction procedure. 
However, as in this paper we focus on the general construction of the concatenation of the GKP and QLDPC codes, and explore the novel benefits of such a concatenation on a more abstract level, we will assume here the use of ideal GKP qubits and perfect operations both during encoding and decoding. That is, we assume here that perfect GKP qubits are generated during the encoding step. Then, all the GKP qubits are subjected to a noisy channel as described above. 
Finally, error correction is performed with perfect ancillas and we assume that no additional noise arises during this step. 
We note that this noise model also provides a good description of practical quantum communication scenarios where the encoding is used to overcome losses in the communication channel.
In such scenarios the dominant source of noise affecting the qubits will be contributed by the channel, while the possible noise due to the use of imperfect ancillas during error correction will be significantly smaller. 
While in a realistic scenario teleportation-based GKP error correction is more robust to the imperfection coming from the use of finitely-squeezed GKP ancillas, under the assumption of ideal error correction the two procedures of Steane as well as teleportation-based error correction become equivalent~\cite{rozpkedek2021quantum,walshe2020continuous,fukui2021all}.

\subsection{GKP Analog Information}
\label{Section_GKP_Analog}

One of the features of the GKP code that makes it very attractive for concatenation as an inner-code with other discrete-variable outer-codes is the generation of the analog information about the GKP data qubits during a measurement of the GKP stabilizers. Specifically, the GKP syndrome $\{Q_0, P_0\} \in [-\sqrt{\pi}/2, \sqrt{\pi}/2)$ not only tells us what displacement needs to be applied to the data qubit in order to correct the errors, but it also provides additional information about the likelihood that the correction displacement will fail to correct the error and hence result in a logical error~\cite{fukui2017analog}. 
Intuitively, if the value of $Q_0$ {(or $P_0$)} is close to the boundary of this interval, e.g., $Q_0 = \sqrt{\pi}/2 - \epsilon$, there is a high likelihood that the actual error displacement is not {$Q_{\text{err}}=Q_0$} but rather $Q_{\text{err}} = Q_0 - \sqrt{\pi}=-\sqrt{\pi}/2 - \epsilon$ so that a feedback displacement by $-Q_0$ will result in a logical $X$ error. On the other hand if $Q_0$ is small, that is e.g., $Q_0 = \epsilon$, then the likelihood of logical error after the feedback displacement is small since the likelihood that the actual error displacement was $Q_{\text{err}} = \pm \sqrt{\pi} + \epsilon$ is small. We can make this error likelihood given the observed syndrome $Q_0$ mathematically precise as it can be expressed as~\cite{Noh-pra20}:
\begin{equation}
\mathbb{P}[\text{logical error} \, | \, Q_0 ; \sigma] = \frac{\sum\limits_{l \in \mathbb{Z}} \exp[\frac{- (Q_0 - (2l+1)\sqrt{\pi})^2}{2\sigma^2}]}{\sum\limits_{l \in \mathbb{Z}} \exp[\frac{- (Q_0 - l\sqrt{\pi})^2}{2\sigma^2}]}.
\label{eq:analogInfoprob}
\end{equation}
Here, $\sigma$ is the standard deviation of the Gaussian random displacement preceding the GKP error correction step. The numerator of the above expression quantifies the likelihood that the syndrome $Q_0$ was generated by measuring a GKP peak which due to the noise was displaced by $l\sqrt{\pi} + Q_0$ for $l$ being an odd integer, hence leading to a logical error. To convert this likelihood into an actual probability of error we include the denominator which acts as a normalisation with respect to the fact that $Q_0$ must correspond to a displacement of a GKP peak by $l\sqrt{\pi} + Q_0$ for some integer $l$, by construction of the GKP code.
The simplest way to use this analog information is to consider post-selection where, depending on the value of $\sigma$, we declare protocol failure if the measured GKP syndrome lies too close to the decision boundary at $\pm \sqrt{\pi}/2$~\cite{Fukui-prx18}. In this way, we can guarantee that if we accept and proceed with the correction feedback displacement, the probability of having a logical error is greatly suppressed. On the other hand, the scenario where the GKP code is concatenated with a higher level discrete-variable outer code offers further possibilities to boost the performance of the outer code~\cite{fukui2017analog,Vuillot-pra19, Noh-pra20,rozpkedek2021quantum}. Specifically, the knowledge of the likelihood of error on each of the GKP qubits after GKP correction provides additional information to the subsequent outer code decoding procedure, which can enable more efficient correction of errors on the higher level.

\subsection{GKP Concatenation Framework}
\begin{figure*}[t]
    \centering
    \includegraphics[width=1.0\textwidth]{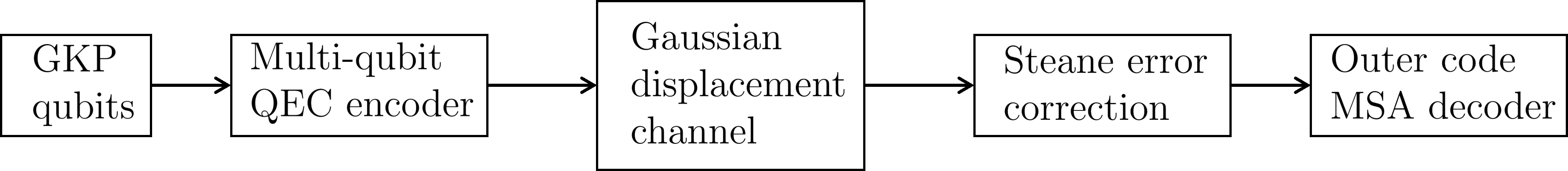}
    \caption{Block diagram representing the concatenation of the GKP code qubits with an outer stabilizer code. The decoder we employ for the outer QLDPC code is the sequential version of the normalized min-sum algorithm (MSA) \cite{HocevarLayered,05CDEFH}. This is a hardware-implementation friendly iterative message passing algorithm that does not sacrifice much in terms of performance. Such an iterative decoder is also able to naturally utilize the analog information from the inner GKP error correction.
    }
    \label{fig:block_diagram_GKP_concatenation}
\end{figure*}

\begin{figure*}[t]
\centering
\includegraphics[width = 0.9\textwidth]{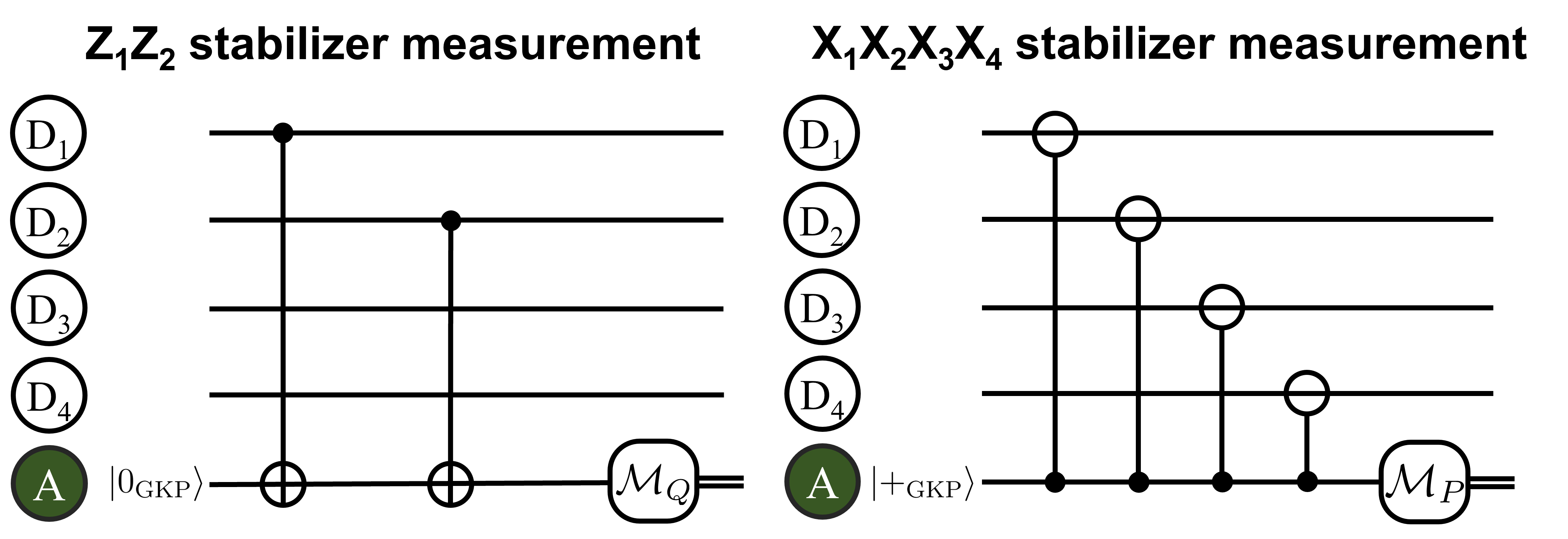}
\caption{Stabilizer measurement for the outer discrete-variable code. Measurement of an outer code stabilizer requires a single ancilla GKP qubit. When measuring a $Z$-stabilizer ($X$-stabilizer), the ancilla is initialized in GKP logical zero (plus) state. After application of the relevant $\text{SUM}$ gates from the data qubits, marked as D, onto the ancilla qubit, marked as A (inverse $\text{SUM}$ gates, $\text{SUM}^\dag$, from the ancilla qubit onto the date qubits), the ancilla is measured in the $Q$ ($P$) quadrature. The one of the two possible outer-code stabilizer values is extracted by checking whether the measured quadrature corresponds to an even or odd multiple of $\sqrt{\pi}$. Figure taken with modifications from~\cite{rozpkedek2021quantum} under the license CC BY 4.0 https://creativecommons.org/licenses/by/4.0/. }
\label{fig:SyndromeOuter}
\end{figure*}

As we have discussed above, the GKP qubit is a two-dimensional subspace of an infinite dimensional Hilbert space that is stabilized by the operators in Eq.~\eqref{eq:GKPstabilizers}.
While the GKP analog information can be used to post-select some of the logical error events, it has been shown that concatenating GKP qubits with an outer stabilizer code provides substantial gains~\cite{Noh-pra20,rozpkedek2021quantum}. Fig.~\ref{fig:block_diagram_GKP_concatenation} demonstrates the general scheme of concatenation used in this work.
In this framework, $n$ GKP qubits are taken as physical qubits for an $\llbr n,k,d \rrbr$ stabilizer code (i.e., an outer code), and hence used to protect $k$ logical qubits with a code distance $d$.
In an error correction cycle, the individual GKP (inner-code) Steane error corrections are performed on the $n$ qubits, and the obtained analog information for each qubit is sent to the decoder for the outer stabilizer code.
The Steane error correction includes syndrome extraction as well as appropriate correction on each GKP qubit.
Then, the outer code stabilizers are measured on the GKP qubits using bare ancilla syndrome measurement circuitry as shown in Fig.~\ref{fig:SyndromeOuter}.
The decoder for this outer code uses information from these syndromes as well as the GKP analog information from the inner GKP code error correction to estimate the most likely Pauli error pattern on the $n$ qubits.
Note that each Pauli element of this error pattern represents the corresponding logical Pauli operator for the GKP code on that GKP qubit.
Hence, the relevant logical Pauli operators are applied on the GKP qubits to fix the outer code syndromes and correct the errors.

As mentioned earlier, in this paper we assume that all GKP ancillas used for syndrome extraction are noiseless, i.e., infinitely squeezed.
Therefore, after applying the GKP-logical Pauli corrections, the qubits together belong to the subspace of the outer code, and each qubit also belongs to the GKP code subspace, either with or without a residual GKP-logical error.

However, we note that if the ancillas used for syndrome extraction are imperfect (i.e., finitely squeezed GKP), then these logical operations need not necessarily bring each GKP qubit to the GKP code space.
This is because, with imperfect ancillas, the result of the first round of GKP corrections does not guarantee that the residual displacement in each quadrature of each GKP qubit is $0$ (no logical error) or $\pm \sqrt{\pi}$ (logical error).
Indeed, the circuit to measure, say, the GKP stabilizer $\hat{S}_Q$ involves the SUM gate from the GKP data qubit to the GKP ancilla as shown in Fig.~\ref{fig:SyndromeInner}. 
The SUM gate from mode 1 to mode 2 is given by $\text{SUM}_{1 \rightarrow 2} = \exp(-i \hat{Q}_1 \hat{P}_2 )$ and it implements an operation under which the quadratures of the two modes transform as follows:
\begin{equation}
\begin{aligned}
\hat{Q}_1 &\rightarrow \hat{Q}_1, \\
\hat{P}_1 &\rightarrow \hat{P}_1 - \hat{P}_2, \\
\hat{Q}_2 &\rightarrow \hat{Q}_1 + \hat{Q}_2, \\
\hat{P}_2 &\rightarrow \hat{P}_2. \\
\end{aligned}
\label{eq:SumGate}
\end{equation}
If the modes 1 and 2 encode GKP qubits, then this transformation implements a CNOT gate. During the measurement of $\hat{S}_Q$, the $Q$-displacement of the data flows through the SUM gate to the ancilla and is read by the measurement (modulo $\sqrt{\pi}$) along with the $Q$-displacement of the ancilla. Moreover, we see that any $P$-displacement of the ancilla will also back-propagate to the data.
As a result, if the ancilla is perfect, then we measure only the data displacement (modulo $\sqrt{\pi}$) and bring it back to the GKP code space (through perfect error correction or a logical error). 
But, if the ancilla is finitely squeezed, then its own erroneous displacement will affect both the measurement as well as the $P$-displacement of the data.
A similar phenomenon occurs while measuring the GKP stabilizer $\hat{S}_P$ as well.
These continuous residual displacements might later accumulate on the ancilla GKP qubits used for outer code stabilizer measurements, effectively leading to logical GKP errors on these ancillas and hence to misidentification of the outer-code stabilizer syndromes~\cite{rozpkedek2021quantum}.

For our simulations, we randomly sample displacement errors according to a chosen noise variance $\sigma^2$, perform GKP Steane error correction for each GKP qubit, determine the residual logical GKP errors, then perform outer code error correction using syndromes and GKP analog information, and find if the estimated Pauli error combined with the residual error pattern after GKP error correction is an outer code stabilizer (no logical error) or an outer code logical operator (logical error).
As we will discuss later, sometimes the iterative decoders that we consider might not be able to find an error pattern that even matches the syndrome.
For the performance plots, we treat such cases as a decoder failure and as causing a logical error (on all the logical qubits).

\section{Quantum LDPC Codes and Syndrome-based Iterative Decoding}
\label{Sec:QLDPC_Code}
In the previous section, we discussed the GKP code and how it can be concatenated with any outer quantum error correction (QEC) code to take advantage of the analog information from GKP stabilizer measurements. As the outer QEC code, we have sparse quantum stabilizer codes --- \emph{quantum low-density parity-check (QLDPC)} codes --- that have been increasingly gaining attention in the past few years.
In this work, we particularly consider QLDPC codes belonging to the \emph{Calderbank-Shor-Steane (CSS)} construction, where the quantum code is completely described by a pair of compatible classical codes~\cite{calderbank1996quantum_exists,Steane-physreva96}.
We demonstrate the concatenation of the GKP (inner) code specifically with QLDPC (outer) codes because QLDPC codes provide several advantages over other families of outer stabilizer codes:
\begin{enumerate}
    
\item Compared to surface codes and color codes of similar code lengths, QLDPC codes boast better code rates with fault-tolerant thresholds supported by low-complexity iterative decoding algorithms \cite{mackay_quantum,gottesman_fault_tolerant_ldpc,HighDimQuantumHPC_Pryadko19, QuantumExpanderCodes_Noisy_syndrome}. 

\item The low weight stabilizers of QLDPC codes, albeit not always spatially local, are favorable for experimental realizations of these codes. 
In the future, when we consider noisy stabilizer measurement circuits, the low weight of these stabilizers will greatly help in controlling error propagation. 

\item The recent discovery of QLDPC codes with rate scaling linearly and distance scaling linearly with the code length~\cite{panteleev2021quantumLinearMinDLocalTestable,panteleev2020quantumLinearMinD,breuckmann2020balanced} can be directly exploited for high rate communications and computing with good reliability.

\item The fast iterative decoders for QLDPC codes, based on the well-known belief propagation (BP) algorithm and its variants \cite{63Gallager}, can naturally exploit the analog information from the inner GKP code discussed in Section \ref{Section_GKP_Analog}.
As we will show in our simulation results, this additional analog information helps the syndrome-based iterative decoders to successfully identify the errors corresponding to the measured syndrome.
\end{enumerate}

In fact, the analog information helps the CSS-QLDPC code and decoder surpass the CSS Hamming bound of $C = 1 - 2 h_2(p)$, where $p$ is the probability that a qubit undergoes an error and $h_2(\cdot)$ denotes the binary entropy function~\cite{calderbank1996quantum_exists,Steane-rspa96,Dennis-jmp02}. 
Given the error rate $p$, the quantity $C$ is the maximum rate achievable by non-degenerate CSS codes that are agnostic to the nature of the underlying physical qubits.
More precisely, we will show a QLDPC code family with rates converging to $R = C$ that can correct errors beyond the error rate $p$ implied by the above expression, using a sequential min-sum decoder \cite{HocevarLayered,Sharon_SerialSchedules} that exploits the GKP analog information.
Here, the error rate $p$ is connected to the variance $\sigma^2$ of the Gaussian random displacement channel through the Eq.~\eqref{eq:p_sigma_relation}. 

\subsection{CSS Codes}

QLDPC codes used in the GKP-concatenation scheme belong to a special case of stabilizer codes called CSS codes. 
In general, stabilizer codes are the quantum analog of classical linear codes~\cite{Gottesman97}. 
A stabilizer group $S$ is a commutative subgroup of the $n$-qubit Pauli group%
\footnote{The $n$-qubit Pauli group $\mathcal{P}_n$ is formed by Kronecker products of $n$ single-qubit Paulis and scalars $\imath^{\kappa}$, where $\kappa \in \mathbb{Z}_4 = \{0,1,2,3\}$.} $\mathcal{P}_n$ that contains only Hermitian Paulis and excludes $-I_n$.
An $\llbracket n,k,d \rrbracket$ quantum stabilizer code $\mathcal{Q}$ encodes $k$ logical qubits into an entangled $n$-qubit code state that is a common $+1$ eigenstate of the elements of $S$. 
The \emph{weight} $w(E)$ of a Pauli operator $E \in \mathcal{P}_n$ is the number of qubits on which it applies a non-identity Pauli matrix.
Any $E \in \mathcal{P}_n \setminus S$ that commutes with all stabilizers has minimum weight $d$, thereby defining the error correction capability of the code. 
The code rate of $\mathcal{Q}$ is defined as $R(\mathcal{Q}) = k/n$.
The $m = n-k$ stabilizer generators of the code describe the rows of the corresponding stabilizer matrix, equivalent to the parity-check matrix. 

The stabilizer group of CSS codes can be generated via purely $X$-type and purely $Z$-type operators given by two compatible classical codes \cite{calderbank1996quantum_exists}.
Consider two classical codes $\MCC_1$ and $\MCC_2$ with parameters $[n,k_1,d_1]$ and $[n,k_2,d_2]$ and parity-check matrices ${\bf H_{\mathrm{X}} }$ and ${\bf H_{\mathrm{Z}} }$, respectively, such that $ \bf H_{\mathrm{X}}  H_{\mathrm{Z}}^T = 0$ (the all-zeros matrix).
Then the CSS code, CSS($\MCC_1, \MCC_2$), induced by these codes is defined by $X$-stabilizer generators obtained by replacing 1s and 0s of ${\bf H_{\mathrm{X}} }$ with $X$s and $I$s, and $Z$-stabilizer generators obtained by replacing 1s and 0s of ${\bf H_{\mathrm{Z}} }$ with $Z$s and $I$s.
Since $ \bf H_{\mathrm{X}}  H_{\mathrm{Z}}^T = 0$, it is guaranteed that the $X$-stabilizers and $Z$-stabilizers commute, thereby forming a valid stabilizer group.
The parameters of CSS($\MCC_1, \MCC_2$) are $\llbr n, k = k_1 + k_2 - n, d \geq \min(d_1,d_2) \rrbr$.
As an example, consider the $\llbr 7,1,3 \rrbr$ Steane code defined by taking $\MCC_1$ and $\MCC_2$ as the classical $[7,4,3]$ Hamming code.
Hence, the $X$- and $Z$-stabilizer generators are given by the same parity-check matrix
\begin{align}
{\bf H_{\mathrm{X}} } = {\bf H_{\mathrm{Z}} } = 
\begin{bmatrix}
1 & 1 & 1 & 0 & 1 & 0 & 0 \\
1 & 1 & 0 & 1 & 0 & 1 & 0 \\
1 & 0 & 1 & 1 & 0 & 0 & 1
\end{bmatrix}.
\end{align}
Thus, the $X$-stabilizer generators are given by $X_1 X_2 X_3 X_5, \ X_1 X_2 X_4 X_6, \ X_1 X_3 X_4 X_7$ and the $Z$-stabilizer generators are given by $Z_1 Z_2 Z_3 Z_5, \ Z_1 Z_2 Z_4 Z_6, \ Z_1 Z_3 Z_4 Z_7$.
Here, $X_i$ (resp. $Z_i$) refers to applying the Pauli $X$ (resp. $Z$) on the $i$-th qubit.

A CSS-based QLDPC code is defined by such a stabilizer generator matrix with low-weight stabilizers. 
In the binary representation of these parity-check matrices, the number of nonzero entries is low compared to the zero entries, giving its name quantum `low-density' parity-check code. 
The sparse parity-check matrix can be represented graphically by a sparse bipartite graph called \emph{Tanner graph} over which the message passing iterative decoders operate. 
We describe the Tanner graphs and message passing decoders in Section \ref{sec:TannerMsgPass}. 

\subsection{Tanner Graph and Message Passing Decoders}
\label{sec:TannerMsgPass}

\begin{figure}
    \centering
    \includegraphics[scale=0.5]{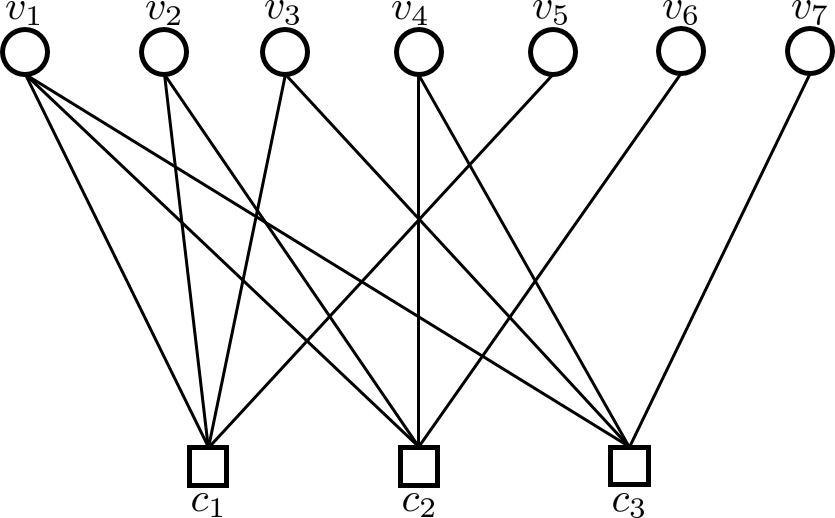}
    \caption{The Tanner graph of the parity-check matrix ${\bf H_{\mathrm{X}} }$ corresponding to the $\llbr 7,1,3 \rrbr$ Steane code.}
    \label{fig:TannerGraphSteaneCode}
\end{figure}

Graphically, the parity-check matrix, $H$, can be represented by a bipartite graph $(G = V \cup C, F)$, where the variable node set $V = \{v_1, \ldots v_n\}$ and check/stabilizer node set $C = \{c_1, \ldots c_m\}$ are connected using the set of edges $F$. The variable nodes and check nodes correspond to the columns and rows of $H$, respectively. 
An edge between a check node $c_i$ and a variable node $v_j$ corresponds to a nonzero entry $H_{i,j}$ in the $i$-th row and $j$-th column of the parity-check matrix. 
In Fig. \ref{fig:TannerGraphSteaneCode}, we show the Tanner graph corresponding to the $X$-checks (or, equivalently, $Z$-checks) of the $\llbr 7,1,3 \rrbr$ Steane code, where we depict the variable nodes as circles and check nodes as squares, and we draw solid lines to represent the edges. 
The rows of the check matrix are indexed by $c_1, c_2, c_3$ and the columns are indexed by $v_1, v_2, \ldots, v_7$.

Message passing decoding algorithms such as BP operate by iteratively passing ``beliefs'', i.e., local posterior probabilities, along the edges of the Tanner graph. 
With its roots in the broad class of Bayesian inference problems, BP is an efficient algorithm to compute marginals of functions on a graphical model \cite{88Pearl,01KFL}. 
Consider a general inference problem where there are $n$ unknowns $\{ x_1, \ldots, x_n \}$ related by a joint function that has a simple factorization into $m$ factors: $f( \{ x_1, \ldots, x_n\}) = \prod_{i=1}^m f_i(\{ x_j; \ j \in V_i \})$, where each $f_i$ is a ``local factor'' that only involves a subset of variables $V_i \subset \{ 1, \ldots, n \}$.
Given some observations of the unknowns (e.g., through a noisy channel), or indirect information about them, the goal of BP is to compute the posterior marginal distribution of each unknown $x_i$, which can then be used to obtain a ``hard'' value for the unknown, e.g., by simple thresholding if the unknowns are Boolean or binary.
Effectively, BP uses the distributivity of addition over multiplication to efficiently compute the marginal 
$$ \sum_{x_1,\ldots, x_{i-1}, x_{i+1}, \ldots, x_n} \prod_{i=1}^m f_i(\{ x_j; \ j \in V_i \}) $$
for the $i$-th unknown $x_i$. In the setting of iterative decoding, BP starts with ``channel information'' about each variable, expressed in terms of an initial distribution, and then passes probability messages via edges on the Tanner graph to eventually compute the posterior marginals.
The factor (or check) nodes in the Tanner graph impose local constraints on the incident variables as dictated by the local factors $f_i$ (which will be stabilizers or parity-checks for our purposes).
If the graph is a tree, then BP exactly computes the posterior marginals and hence performs maximum-a-posteriori (MAP) inference.
If the graph has loops, then BP can still perform very well, although not optimal, if certain troublesome graph configurations are avoided.
The main advantage in such scenarios is the low complexity of BP compared to MAP.
In the next section, we provide a more specific discussion for the iterative decoding of QLDPC codes.

\subsection{Iterative Decoding of QLDPC Codes}

For classical LDPC codes $\mathcal{C}$, decoders observe the received vector $\underline{y} = [y_1,y_2 \ldots, y_{n}]$, which is typically the transmitted codeword $\underline{x} = [x_1, x_2,\ldots, x_{n}] \in \mathcal{C}$ added with channel noise, to infer the transmitted codeword $\underline{x}$.
The bitwise channel log-likelihood ratios (LLRs) that are calculated as $\lambda_i = \ln \frac{\Pr(y_i|x_i = 0)}{\Pr(y_i|x_i = 1)}$ are used for the initialization of the iterative BP algorithm. 
For a standard classical channel called the additive white Gaussian noise (AWGN) channel, we have $y_i = x_i + n_i$ with $x_i \in \{ \pm 1 \} (0 \mapsto +1,\ 1 \mapsto -1)$ and $n_i$ being a standard Gaussian random variable with mean $0$ and variance $\nu^2$.
Therefore, the LLRs for the AWGN channel are as follows: 
\begin{equation*}
    \begin{aligned}
    \lambda_i = \ln \frac{\Pr(y_i|x_i = +1)}{\Pr(y_i|x_i = -1)} = \ln \frac{\exp\left( - \frac{(y_i - 1)^2}{2\nu^2} \right)}{\exp\left( - \frac{(y_i + 1)^2}{2\nu^2} \right)} = \frac{2 y_i}{\nu^2}.
    \end{aligned}
\end{equation*}
Starting with the LLRs as initial (channel) messages from variable nodes (representing codeword bits), these decoders iteratively pass messages between variable nodes and check nodes (representing parity-checks) in the Tanner graph to determine the \emph{a posteriori} value for each bit of the codeword.

One of the key differences in decoding QLDPC codes is that there is no classical notion of a received vector. 
Instead, measurement of the stabilizer generators indicates if an error acted upon the quantum state or not. 
For CSS codes, measurement of stabilizer generators, i.e., the rows of the parity-check matrices ${\bf H_{\mathrm{X}} }$ and ${\bf H_{\mathrm{Z}} }$, yields the syndrome ${\bf s} = [{\bf s_X}, {\bf s_Z}]$ corresponding to the error vector $\bf{e} =[ \bf{e_Z}, \bf{e_X}]$, where ${\bf s_X} = {\bf H_{\mathrm{X}} } {\bf e_Z}^T$ and ${\bf s_Z} =  {\bf H_{\mathrm{Z}} } {\bf e_X}^T$. 

The decoder's task is to identify an error vector that corresponds to the measured syndrome ${\bf s}$. 
Hence, the iterative decoder used for classical LDPC codes is modified to address this syndrome-based decoding scenario encountered in QEC. 
We can perform decoding for the $X$
and $Z$ errors separately as our noise model only produces independent $X$ and $Z$ errors. For simplicity of explaining the syndrome-based decoder, we use $\bf H, \bf s,$ and  $\bf e$ for the parity-check matrix, measured syndrome, and the error vector, respectively.

Next, we describe a syndrome-based version of the iterative message passing algorithm used in the decoding of QLDPC codes. 
Suppose we have an (unknown) binary error vector $\bf {e} = (e_1, \ldots, e_n)$ which resulted in a measured syndrome $\bf s$.
In this setting, the variable nodes of the Tanner graph represent error bits rather than codeword bits, and the check nodes are initialized with the respective binary measured syndrome bits. 
The objective of syndrome-based iterative message passing decoders is also to iteratively compute the \emph{a posteriori} probabilities, but of the \emph{error} bits conditioned on the value of the measured syndrome, $\Pr(e_i|\bf s)$ for $i \in \{1,\ldots,n\}$. 
This computation relies on the initialization of the beliefs/messages, followed by passing them over the Tanner graph following the decoder update rules. 
Conventionally, the initialization step of QLDPC decoders does not exploit any analog information and hence, the initial messages that are passed are equal for all variable nodes. This step, which is the  crucial difference in our QLDPC-GKP coding scheme, is explained next.

In syndrome-based iterative decoders, the initial belief of $e_i$ being $0$ rather than $1$ is defined to be the LLR, i.e., the beliefs are initialized with the same LLR value ($\lambda$) for all variable nodes, 
\begin{align}
    \lambda_i = \ln \frac{\Pr(e_i= 0)}{\Pr(e_i = 1)}.
    \label{eq:LLR_syndDef}
\end{align}
Intuitively, this initialization step can be interpreted as the decoder assuming no error on any qubit, and proceeding to find the correct error pattern that matches the syndrome.
Note that this method of initializing the LLR (of the error reliability) will be used only when there is no  GKP analog information available for the iterative decoder at the variable nodes.

For our QLDPC-GKP code, without GKP syndrome $\{Q_0, P_0\}$ and GKP analog information, we encounter the \emph{same} initial LLR for all variable nodes. 
With this idea, it will be equivalent if we set $Q_0$ to zero,  corresponding to no error on any qubits, and use Eq.~\eqref{eq:analogInfoprob} to calculate the LLR as defined in Eq.~\eqref{eq:LLR_syndDef}. 
Hence, initialization for decoding without the GKP analog information is given as 
\begin{equation}
\begin{aligned}
 \lambda_i & = \ln \frac{\mathbb{P}[\text{No logical error}]}{\mathbb{P}[\text{Logical error}]} \\ & = \ln \frac{\sum_{l \in \mathbb{Z}} \exp[-(2l\sqrt{\pi})^2/(2\sigma^2)]}{\sum_{l \in \mathbb{Z}} \exp[-((2l+1)\sqrt{\pi})^2/(2\sigma^2)]} \\ & \approx  -\ln{2} + \frac{\pi}{2\sigma^2} \ (\text{if}\ \sigma \ll 1).
\label{eq:NoAnalogInfoprob}
\end{aligned}
\end{equation}
However, for ease of decoder implementation, it is common to set these LLRs to a small positive channel value such as $+1$, to slightly bias it towards zero error pattern as initialization~\cite{francisco_Access21_quantum_MSA_fpga}. Our simulation experiments confirmed that this constant value on all variable nodes has very similar decoding performance as using~\eqref{eq:NoAnalogInfoprob} for initialization.

\begin{figure}[t]     
\centering
\begin{subfigure}[b]{0.3\textwidth} \centering \includegraphics[width=\textwidth]{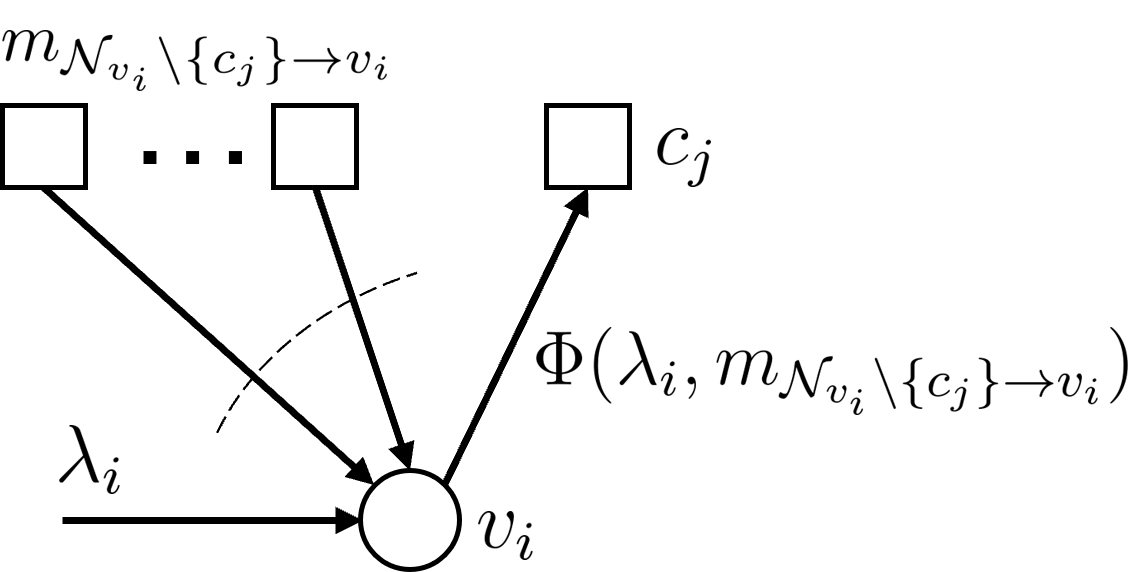} \caption{Variable node update} \label{fig:VNU} \end{subfigure}
\begin{subfigure}[b]{0.3\textwidth} \centering \includegraphics[width=\textwidth]{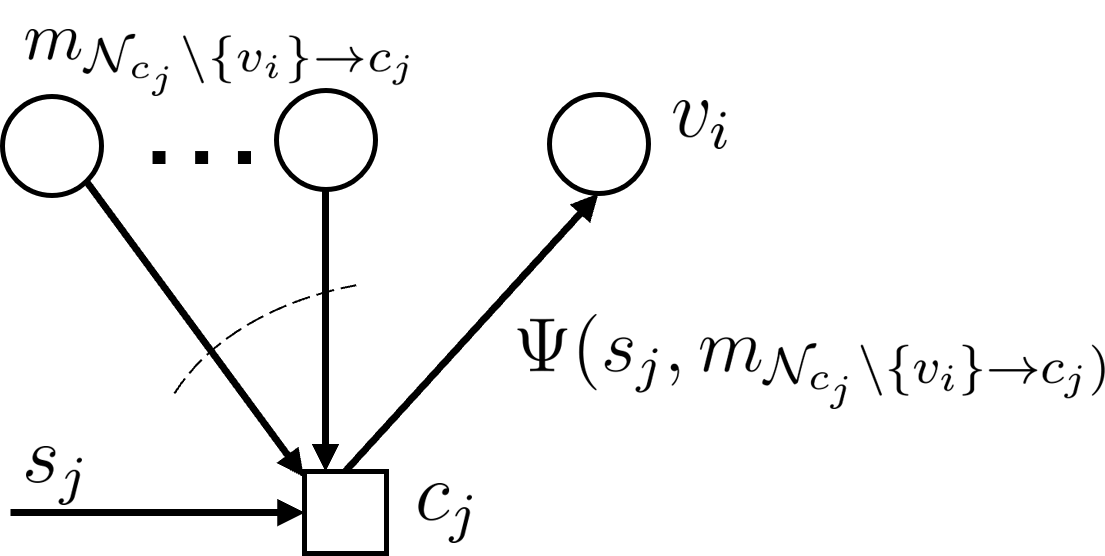} \caption{Check node update} \label{fig:CNU} \end{subfigure}
\caption{ Computation of the variable node message uses VNU function $\Phi$, and that of the check node message uses CNU function $\Psi$. Each message passing iteration comprises of these two message update functions.
}
\label{fig:UpdateRules}
\end{figure}

There are two types of functions in an iterative message passing decoder, namely, the variable node update (VNU) function $\Phi$ and the check node update (CNU) function $\Psi$ as depicted in Fig. \ref{fig:UpdateRules}.
The $\Phi$ (respectively, $\Psi$) function computes the messages propagating from variable (respectively, check) nodes to check (respectively, variable) nodes. 
These decoders pass the messages/beliefs along the edges of the Tanner graph in an iterative fashion. 
Let the set of neighbors of a node $x$ be denoted as $\mathcal{N}_x$, and let the cardinality $|\mathcal{N}_x|$ indicate the node degree $d_x$. 
Let the message passed from node $x$ to node $y$ in the $k$-th iteration be denoted as $m_{x \to y}^{(k)}$. 
Both the check node and variable node update functions use the ``extrinsic'' message passing principle, wherein a message from node $x$ to $y$ is computed as some function of all incoming \emph{extrinsic messages}, i.e., from $\mathcal{N}_x \backslash \{ y \}$ --- all neighbors of the node $x$ except $y$. 
\begin{enumerate}
    \item Variable node update: The VNU function $\Phi$ at each variable node is computed using the corresponding LLR value and all extrinsic incoming check node messages (See Fig.~\ref{fig:VNU}).
    
    For every variable node $v_i \in V$, 
    \begin{equation}
    m_{v_i \to c_j} = \Phi(\lambda_i, m_{c\to v_i}),
    \label{eq:VNU}
    \end{equation} 
    where $c \in \mathcal{N}_{v_i} \backslash \{ c_j \}$ and $c_j \in \mathcal{N}_{v_i}$.
    \item Check node update: The CNU function $\Psi$ at each check node is computed using the corresponding syndrome value and all extrinsic incoming variable node messages (See Fig.~\ref{fig:CNU}).

    For every check node $c_j \in C$, 
    \begin{equation}
    m_{c_j \to v_i} = \Psi(s_j, m_{v\to c_j}), 
    \label{eq:CNU}
    \end{equation}
    where $v \in \mathcal{N}_{c_j} \backslash \{ v_i \}$ and $v_i \in \mathcal{N}_{c_j}$.
\end{enumerate}

\begin{figure*}
    \centering
    \includegraphics[scale=0.65]{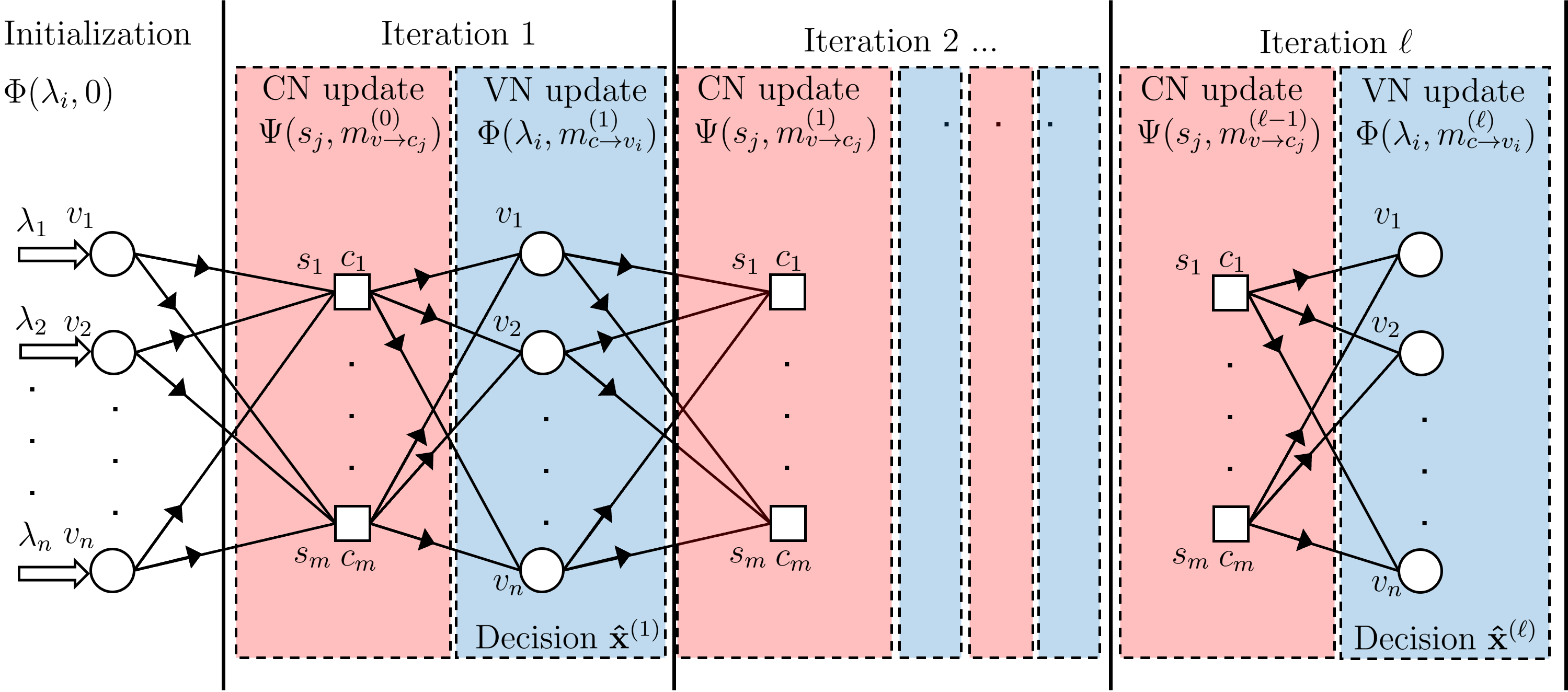}
    \caption{The Tanner graph is unrolled showing the update rules at each iteration of the syndrome-based iterative message passing decoder. The iterative decoder's task is to infer an error vector whose syndrome is equal to the input syndrome, $\mathbf{s}$, using the message passing update rules over iterations. In each iteration step, the check node update $\Psi$ is followed by the variable node update $\Phi$ and the decision update $\hat{\Phi}$. The decoding procedure is initialized with analog log-likelihood ratios (LLRs) corresponding to each of the inner-GKP error likelihoods for the variable nodes, unlike traditional syndrome based decoders that are initialized by LLR corresponding to the all-zero error pattern. The iterative decoding procedure is successful, and outputs the error vector estimate $\mathbf{\hat{x}}$, if the output syndrome $\mathbf{\hat{s}}$ matches with the input syndrome $\mathbf{s}$. Even in such a ``success'' event the decoder could have miscorrected the error. The iterative decoder is said to have failed to converge if the estimated error pattern does not correspond to the input syndrome even after the preset maximum number of iterations, $\ell_{\max}$.}
    \label{fig:iterative_decoding}
\end{figure*}

A message passing decoder employs these two update functions in each iteration $\ell \le \ell_{\max}$ and then determines the error on each variable node. 
At iteration $\ell$, the error estimate $\hat{x}_i^{(\ell)}$ is based on the sign of the \emph{decision update function} output $\hat{\Phi}$ that uses the LLR value and \emph{all} incoming check node messages to the variable node $v_i$, i.e., $\hat{\Phi}(\lambda_i, m_{c\to v_i}^{(\ell)})$, where $c \in \mathcal{N}_{v_i}$. The error estimate is $\hat{x}_i^{(\ell)} = (1 - \text{sgn} (\hat{\Phi}(\lambda_i, m_{c\to v_i}^{(\ell)})) )/2$ where the sign function is defined as $\text{sgn}(A) = -1$ if $A<0$, and $+1$ otherwise. 
The output of the decoder at $\ell$-th iteration, denoted by $\mathbf{\hat{x}}^{(\ell)}=(\hat{x}^{(\ell)}_1,\hat{x}^{(\ell)}_2,\ldots,\hat{x}^{(\ell)}_n)$, is used to check whether all parity-check equations are matched, i.e., the syndrome at $\ell^{\text{th}}$ iteration ${\bf \hat{s}^{(\ell)}} = {\bf \hat{x}^{(\ell)}}\cdot {\bf H}^{\rm T}$ is equal to the input syndrome ${\bf s}$, in which case iterative decoding terminates and outputs $\bf \hat{x}^{(\ell)}$ as the error vector.
Otherwise, the iterative decoding steps continue until a predefined maximum number of iterations, denoted by $\ell_{\max}$, is reached.
A schematic of the iterative message passing decoding procedure is shown in Fig.~\ref{fig:iterative_decoding}.

Typically, a variant of the BP decoder known as a \emph{min-sum algorithm} (MSA) is used in practice due to its low complexity and hardware-friendly nature \cite{HocevarLayered}. As the name `min-sum' suggests, the CNU function is a minimum computation and the VNU function is a summation of the incoming messages as follows: 
\begin{equation}
\begin{aligned} 
\label{eq:CNMSA}
\qquad \Psi(s_j, m_{v\to c_j}) =  \textrm{sgn}(s_j) \cdot & \prod_{v \in \mathcal{N}_{c_j} \backslash \{ v_i \}}\textrm{sgn}(m_{v\to c_j})\cdot \\ &  \underset{v \in \mathcal{N}_{c_j}\backslash \{ v_i \}}{\min}\left | m_{v\to c_j} \right |,
\end{aligned}
\end{equation}
\begin{align} 
\label{eq:VNMSA}
\Phi(\lambda_i, m_{c\to v_i}) &= \lambda_i + \sum_{c \in \mathcal{N}_{v_i} \backslash \{ c_j \} }m_{c\to v_i}.
\end{align}

Min-sum update rules are less complex than traditional BP decoders, but they suffer from performance degradation which is improved typically using a normalized MSA \cite{05CDEFH} wherein check node messages are multiplied (in Eq. \eqref{eq:CNMSA}) by a scalar $\beta, (0 < \beta < 1)$ as a correction factor.

In addition to the update functions $\Psi$ and $\Phi$, the order in which the check node and variable node messages are updated is also important for decoder implementation. Commonly used scheduling strategies are parallel/flooding and sequential/layered schedules. 
The decoder is said to follow a parallel schedule if, in an iteration step, all CNs are updated simultaneously and subsequently all VNs are updated simultaneously. 
In contrast, a sequential schedule updates the messages sequentially, or one-by-one, until all check nodes and variable nodes are updated in an iteration step, e.g., the sequential update with a column update order - $v_1$ followed by $v_2, \ldots, v_n$. 
In this paper, we present simulation results using the concatenated GKP framework where the outer code decoders follow the normalized min-sum algorithm with sequential updating schedules.  
We now describe how the syndrome-based MSA is applied in our concatenated GKP framework.

\subsection{Simulation Setup and Using the GKP Analog Information in the MSA}

The MSA decoder can be used for decoding QLDPC codes by using the binary parity-check matrix representing the stabilizers to define the corresponding Tanner graph.
Since we consider CSS codes, which are defined by two parity-check matrices ${\bf H_X }$ and ${\bf H_Z}$, and our noise model only produces independent $X$ and $Z$ errors, we employ the MSA algorithm separately on their Tanner graphs.

First, we sample random displacements in the $Q$ and $P$ quadratures for each GKP qubit, according to the noise variance $\sigma^2$.
Then, we simulate Steane error correction on each GKP qubit according to the same procedure in~\cite{Gottesman-pra01} (see~\cite[Supplementary Information]{rozpkedek2021quantum} for more details).
This provides the residual displacement for each qubit after correction, which will be $0$ or $\pm \sqrt{\pi}$ (or integer multiple of $\pm \sqrt{\pi}$) since the ancillas are taken to be perfect (infinitely squeezed).
We convert this into a sequence of (GKP-logical) Pauli operators acting on the GKP qubits, which forms the true error for the outer code.

Also, based on the GKP measurements, we calculate the probability of GKP logical $X$ error, $p_{X_i}$, and logical $Z$ error, $p_{Z_i}$, independently according to Eq. \eqref{eq:analogInfoprob}, which constitutes the analog information for the $i^{\text{th}}$ qubit.
These values (for all $n$ GKP qubits) are then passed on as channel inputs to the syndrome-based MSA decoder for the outer code.
Subsequently, we simulate syndrome extraction for the stabilizer generators of the outer QLDPC code and calculate the input syndrome $\mathbf{s}$, again according to the same procedure in~\cite{Gottesman-pra01,rozpkedek2021quantum}.
The MSA decoder for $X$ errors (resp. $Z$ errors) is initialized by setting the LLR for each qubit to be $\ln \frac{1 - p_{X_i}}{p_{X_i}}$ (resp. $\ln \frac{1 - p_{Z_i}}{p_{Z_i}}$).
Note that in this LLR calculation, we utilize the GKP syndrome $\{Q_0, P_0\}$, therefore the LLR is unique to each qubit. 
This is how the GKP analog information is used by the outer code decoder.
With the aforesaid initialization, the MSA decoder is run until it obtains an error matching the syndrome or until it reaches the maximum number of iterations, $\ell_{\max}$, whichever occurs earlier. 
In the scenario where it does find a syndrome-matching error pattern, we multiply it with the true error computed earlier to see if the result is a stabilizer or logical operator (of the outer code).
If it is the former, then error correction worked perfectly, but if it is the latter, then we count that as a logical error.
In the scenario where the MSA reaches $\ell_{\max}$ and is unable to find a syndrome-matching error pattern, we also declare a logical error for simulation purposes.

It is interesting to note that, in all of our simulations, whenever the decoder found a syndrome-matching error pattern, it corrects the error perfectly with high probability.
Therefore, almost all logical error events in the performance curves correspond to reaching $\ell_{\max}$ and not finding a matching error pattern.

\section{Simulation Results and Discussion}
\label{sec:SimResults}

In this section, we demonstrate the improved threshold and advantages of our QLDPC-GKP concatenation scheme. For the simulations, we choose the recently developed family of lifted product CSS QLDPC codes as the outer code.
First, we describe this code construction in general and then use specific examples to illustrate the advantages of the QLDPC-GKP scheme.

\subsection{Lifted Product QLDPC codes} \label{sec:LPQLDPC}

We consider the lifted product (LP) QLDPC codes as the outer code. LP codes proposed by Panteleev and Kalachev \cite{panteleev2020quantumLinearMinD} are \emph{lifted} versions of hypergraph product codes \cite{hypergraphProductCodeTillich,hypergraphProductCodekovalev_pryadko_improved} and this family has a nonzero asymptotic rate with an almost linear distance scaling under increasing code length. 
Furthermore, the LP-QLDPC codes enable us to create finite length QLDPC codes from good classical (and quantum) \emph{quasi-cyclic (QC)} LDPC codes. 
In classical error correction schemes, QC-LDPC codes are widely used as they provide flexibility both in terms of code construction and decoder implementation, compared to random LDPC code constructions.
We choose QC-LDPC codes as constituent classical LDPC codes to construct LP code families for the QLDPC-GKP concatenation scheme. 

In \cite{Fossorier04}, Fossorier introduced the QC-LDPC codes whose parity-check matrices are obtained by expanding a base matrix $B$ of size $m_b\times n_b$. Each entry of the base matrix is expanded to a binary square matrix of size $L \times L$. 
A non-negative entry $b$ in the base matrix is replaced by a cyclic permutation matrix (${\bf CPM}_{L}(b)$), i.e., binary $L \times L$ identity matrix cyclically right-shifted by $b$ columns. 
Negative entries in $B$ are replaced by $L \times L$ zero matrices in the parity-check matrix. 
We refer to $L$ as the circulant size or lift size of the QC-LDPC code. Hence, the QC-LDPC code's parity-check matrix $H$ is concisely expressed in terms of its base matrix $B$ --- as an $m_b\times n_b$ array of circulant shifts, and the circulant size $L$. 
If negative entries are avoided in $B$, then the corresponding QC-LDPC code is referred to as $(m_b,n_b)$-regular QC-LDPC code since each column in $H$ has weight $m_b$ and each row has weight $n_b$.

\begin{example}
\normalfont
Consider the $[155, 64, 20]$ code $\mathcal{C}$, a $(3,5)$-regular QC-LDPC code~\cite{TannerCode} of circulant size $L = 31$, where the base matrix of $B$ is defined as follows: 
\begin{align}
B & = 
\begin{bmatrix}
 1 &  2 &  4 &  8 & 16 \\
 5 & 10 & 20 &  9 & 18 \\
25 & 19 &  7 & 14 & 28
\end{bmatrix}.
\label{eq_QCHmatrix}
\end{align}
In binary representation, the parity-check matrix $H$ is a $93 \times 155$ matrix, but it only has rank $91$, which is why the code encodes $155 - 91 = 64$ bits.
Hence, this $(3,5)$-regular classical Tanner code~\cite{TannerCode} is constructed using CPMs and has relatively large minimum distance of $20$. 

For the LP construction, we also need the conjugate transpose of the base matrix representation, $B^*$. 
For the Tanner code above, 
\begin{align}
{B^*} = 
\begin{bmatrix}
30  &  26  &   6\\
29  &  21  &  12\\
27  &  11  &  24\\
23  &  22  &  17\\
15  &  13  &   3
\end{bmatrix}.
\end{align}
The asterisk in the superscript of $B^*$ is a conjugate transpose over the matrix ring as in \cite{panteleev2020quantumLinearMinD}, so that the binary representation of $B^*$ is exactly the matrix transpose of the parity-check matrix $H$. 
\end{example}

We construct the $\llbracket n, k, d \rrbracket$ LP code \cite{panteleev2020quantumLinearMinD} by performing the Kronecker product of the base matrix $B$ and its conjugate transpose $B^*$ of the QC-LDPC code as follows. 
The lifted product construction $LP(B, B^*)$ gives stabilizer parity-check matrices $\bf H_{\mathrm{X}}$ and $\bf H_{\mathrm{Z}}$ whose base matrices are 
\begin{equation}
\label{eq:LP_QLDPC_eq}
\begin{aligned}
&{\bf B_{\mathrm{X}} } = 
\begin{bmatrix}
{B } \otimes { I_{n_b} } \, , & {  I_{m_b} } \otimes {B^* }
\end{bmatrix}\\
& \qquad \qquad \text{and} \\
&{\bf B_{\mathrm{Z}} } = 
\begin{bmatrix}
{I_{n_b} } \otimes {B } \, , & {B^* } \otimes { I_{m_b} }
\end{bmatrix},
\end{aligned}
\end{equation}

respectively. The size and properties of the base matrix $B$ determine the properties of the LP code. The code length of $LP(B, B^*)$ is $n = \ell(n_b^2 + m_b^2)$, code dimension $k \ge \ell(n_b - m_b)^2$, and hence, the rate of such ensembles is $r \ge \dfrac{(n_b-m_b)^2}{(n_b^2 + m_b^2)}$. For example, from the $[155, 64, 20]$ Tanner code with base matrix $B$ of size $3 \times 5$ and circulant size $L=31$, we obtain the $\llbracket1054, 140, 20\rrbracket$ LP code. 

For the QLDPC-GKP concatenation scheme, we construct LP-QLDPC code families from $(m_b,n_b)$ families of QC-LDPC codes with different parameters.
Tables~\ref{table:CodeParams34} and~\ref{table:CodeParams35} respectively provide parameters of the LP codes constructed from $(3,4)$-regular and $(3,5)$-regular QC-LDPC codes. 
The LP code families in Tables~\ref{table:CodeParams34} and~\ref{table:CodeParams35} are of asymptotic rates $0.04$ and $0.118$, respectively. 
The last columns in Tables~\ref{table:CodeParams34} and~\ref{table:CodeParams35} indicate the \emph{girth}, which is the length of the shortest cycle in the Tanner graph.  

\begin{table}
	\centering
	\begin{tabular}{|c|c|c|c|c|c|}
	\hline 
		LP-QLDPC & Rate & $L$ & Girth \\
		\hline
		$\llbr175,19,\le10\rrbr$ & 0.109 & 7 & 6 \\
		\hline
		$\llbr225,21,\le12\rrbr$ & 0.093 & 9 & 6 \\
		\hline
		$\llbr425,29,\le18\rrbr$ & 0.068 & 17 & 8 \\
		\hline
		$\llbr475,31,\le20\rrbr$ & 0.065 & 19 & 8 \\
		\hline
	\end{tabular}
	\caption{\label{table:CodeParams34} Lifted product codes of LP04 (with asymptotic rate $0.04$) family from $(3,4)$-regular QC-LDPC codes.}
\end{table}

\begin{table}
	\centering
	\begin{tabular}{|c|c|c|c|c|c|}
	\hline 
		LP-QLDPC & Rate & $L$ &  Girth \\
		\hline
		$\llbr544,80,\le12\rrbr$ & 0.147 & 16 & 8 \\
		\hline
		$\llbr714,100,\le16\rrbr$ & 0.140 & 21 & 8 \\
		\hline
		$\llbr1020,136,\le20\rrbr$ & 0.133 & 30 & 8 \\
		\hline
	\end{tabular}
	\caption{\label{table:CodeParams35} Lifted product codes of LP118 (with asymptotic rate $0.118$) family from $(3,5)$-regular QC-LDPC codes}
\end{table}

\subsection{Threshold Plots}
In the following simulations, we decode the outer QLDPC codes using a syndrome-based MSA (normalization factor $\beta$ is set to $0.75$ empirically) with a sequential schedule for a preset maximum of $\ell_{\max} = 100$ iterations. 
The logical error rates for codes in both LP04 and LP118 families are plotted as a function of $\sigma$, the standard deviation of the Gaussian random displacement preceding the GKP error correction step.
For plotting the threshold curves, for each noise variance $\sigma^2$ spaced at 0.01 we collect at least 10,000 logical errors (sufficient to avoid statistical errors). 

In the simulation plots, we observe a transition from error suppression to error enhancement with increasing $\sigma$, signifying the existence of the error threshold. Below the threshold, the logical error rate decreases with increasing code length leading to a suppression of the logical error rate, whereas increasing the code length leads to an increased logical error rate above the threshold value. We characterize the error thresholds for the following two cases: one where the outer code decoder makes use of the analog information obtained from the inner GKP error correction protocol (decoder w/ analog information) and the other where the decoder initialization does not use the analog information (decoder w/o analog information).

\begin{figure*}[t]     
\centering
\begin{subfigure}[b]{0.49\textwidth} \centering \includegraphics[trim=30 0 60 15,clip, width=\textwidth]{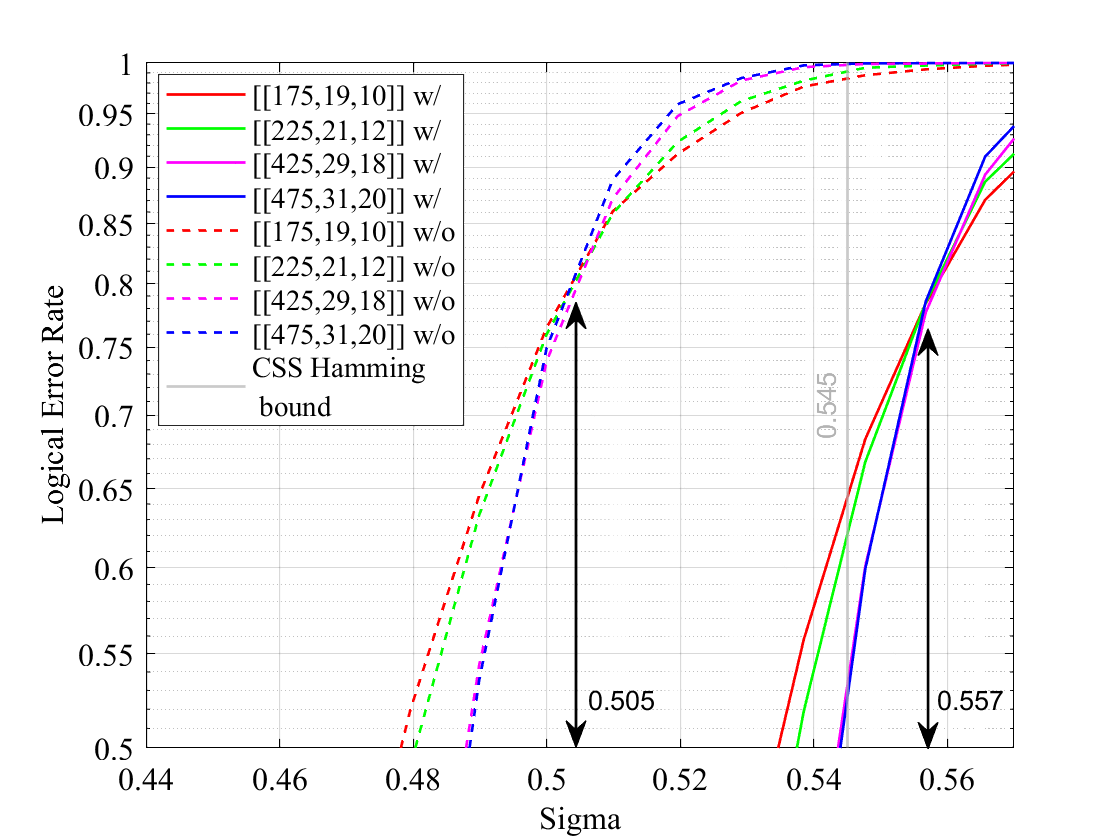} \caption{LP04 code family} \label{fig:LP04} \end{subfigure}
\hspace{1mm}
\begin{subfigure}[b]{0.49\textwidth} \centering \includegraphics[trim=30 0 58 15,clip, width=\textwidth]{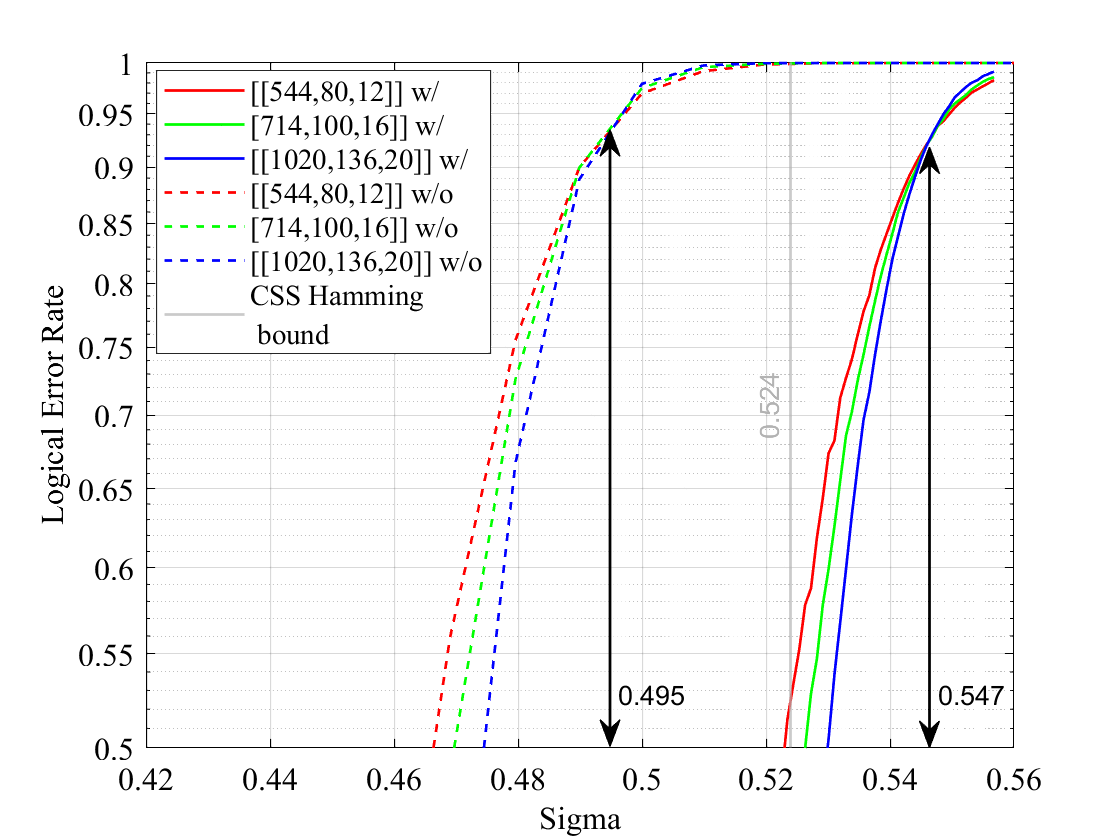} \caption{LP118 code family} \label{fig:LP118} \end{subfigure}
\caption{The set of curves in figures (a) and (b) correspond to the GKP code concatenated with LP-QLDPC codes obtained from lifting regular $(3,4)$ and $(3,5)$ QC-LDPC codes, respectively, of increasing code length (and distance). The dashed curves correspond to the sequential MSA decoder without using the analog information (from the inner GKP error correction), whereas the solid curves make use of the analog information for decoding the outer code. 
The transition of the curves with increasing $\sigma$ signifies an error threshold.
The threshold of the LP04 code family is improved from $\sigma = 0.505$ to $\sigma = 0.557$ with the help of GKP analog information. 
Similarly, the threshold of the LP118 code family is improved from $\sigma = 0.495$ to $\sigma = 0.547$. Furthermore, the improved thresholds for the LP04 and LP118 code families surpass their respective CSS Hamming bounds, highlighted as gray vertical lines at $\sigma = 0.545$ and $\sigma = 0.524$ in the plots.}
\label{fig:threshold_dv3dc4_5_with_withoutSoftInfoLayered}
\end{figure*}

In Fig. \ref{fig:threshold_dv3dc4_5_with_withoutSoftInfoLayered}, we observe the error thresholds for these two cases of outer code decoding for the two LP code families in Tables~\ref{table:CodeParams34} and~\ref{table:CodeParams35}. 
For LP04, the ensemble of LP codes with an asymptotic rate $1/25 = 0.04$, the threshold for the decoder using analog information is observed at around $\sigma = 0.557$. This is better than the threshold of $0.505$ for the same decoder without utilizing the analog information. 
Similarly, for the LP118, the ensemble of LP codes with an asymptotic rate $4/34 \approx 0.118$, the threshold for the decoder using analog information is observed at around $\sigma = 0.547$, in comparison to a lower threshold of $0.495$ for the same decoder without utilizing the analog information.

\subsection{Discussion on Capacity}
\label{sec:capacity_discussion}
In the concatenated coding scheme proposed in this paper, we construct a sequence of CSS LP-QLDPC codes whose asymptotic rate is $R = C(p)$, where $C(p)$ denotes the CSS Hamming bound discussed earlier and $p$ is related to the variance of the Gaussian random displacement channel through the expression in Eq.~\eqref{eq:p_sigma_relation}.
Then, using simulations, we show that our scheme can surpass the threshold $p$ (or, equivalently, $\sigma$) implied by the above bound.
Quantitatively, for the LP04 family, if we set $C(p) = 1/25 = 0.04$, then the CSS Hamming bound implies a threshold of $p = 0.104$, or equivalently $\sigma = 0.545$.
However, by a combination of the two factors stated below, we observe a threshold of $\sigma = 0.557$ in Fig. \ref{fig:threshold_dv3dc4_5_with_withoutSoftInfoLayered}.
Similarly, the CSS Hamming bound implies a threshold of $\sigma = 0.524$ for the LP118 family, whereas the two factors below lead to a threshold of $\sigma = 0.547$ as seen in Fig. \ref{fig:threshold_dv3dc4_5_with_withoutSoftInfoLayered}.

This provides another strong indication of the utility of the concatenation with GKP codes.
In particular, the above result of surpassing the threshold of the CSS Hamming bound is made possible through two factors:
(a) the use of GKP analog information in the MSA decoder, and
(b) the sequential update schedule for the MSA decoder.
We verified that the above result breaks if either of these choices is dropped.
The sequential schedule has recently been shown to outperform the more common parallel/flooding schedule~\cite{refinedBP_QLDPC_2020,Raveendran-Quantum2021}.
However, since these LP-QLDPC codes are degenerate, it is unclear how much this degeneracy contributes to surpassing the CSS Hamming bound.
There has been a negative result for qudits of dimension at least $5$~\cite{Sarvepalli-pra10}, where the authors show that such qudit CSS codes cannot surpass the bound.
For qubits, we leave this important question for future investigation.

\subsection{Effect of GKP Analog Information on Error Floor}
\label{sec:Errorfloor}
\begin{figure}[t]
    \centering
    \includegraphics[trim=30 0 60 15,clip,width=0.49\textwidth]{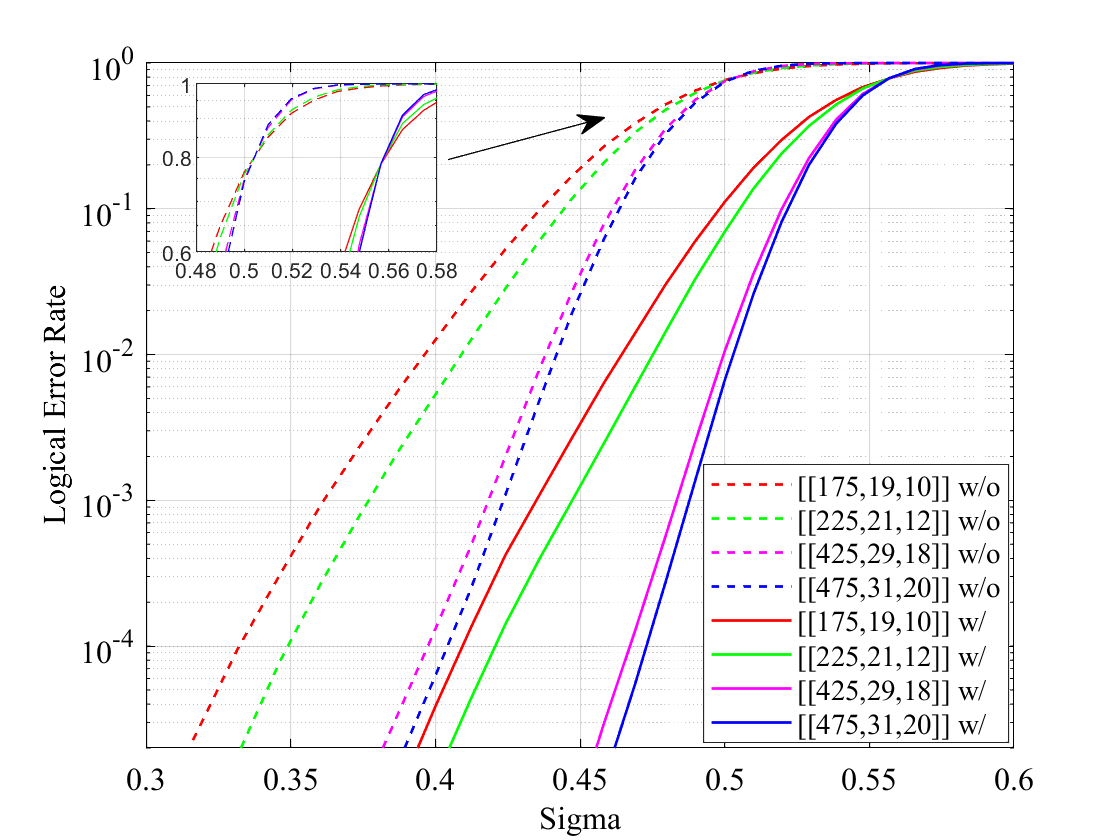}
    \caption{The set of performance curves corresponding to the LP04 family, simulated more ``deeply'' for low $\sigma$ values. The dashed curves correspond to the standard sequential MSA decoder without using the analog information (from the inner GKP error correction), whereas the solid curves make use of the the analog information for decoding the outer code. The threshold of the code family is improved from $\sigma = 0.505$  to $0.557$. This is also above the theoretical CSS Hamming bound that implies a threshold of $\sigma = 0.545$ for an asymptotic rate $(1/25)$, demonstrating the advantage of GKP concatenation.}
    \label{fig:threshold_dv3dc4_with_withoutSoftInfoLayered}
\end{figure}

\begin{figure}[t]
    \centering
    \includegraphics[trim=30 0 60 15,clip,width=0.49\textwidth]{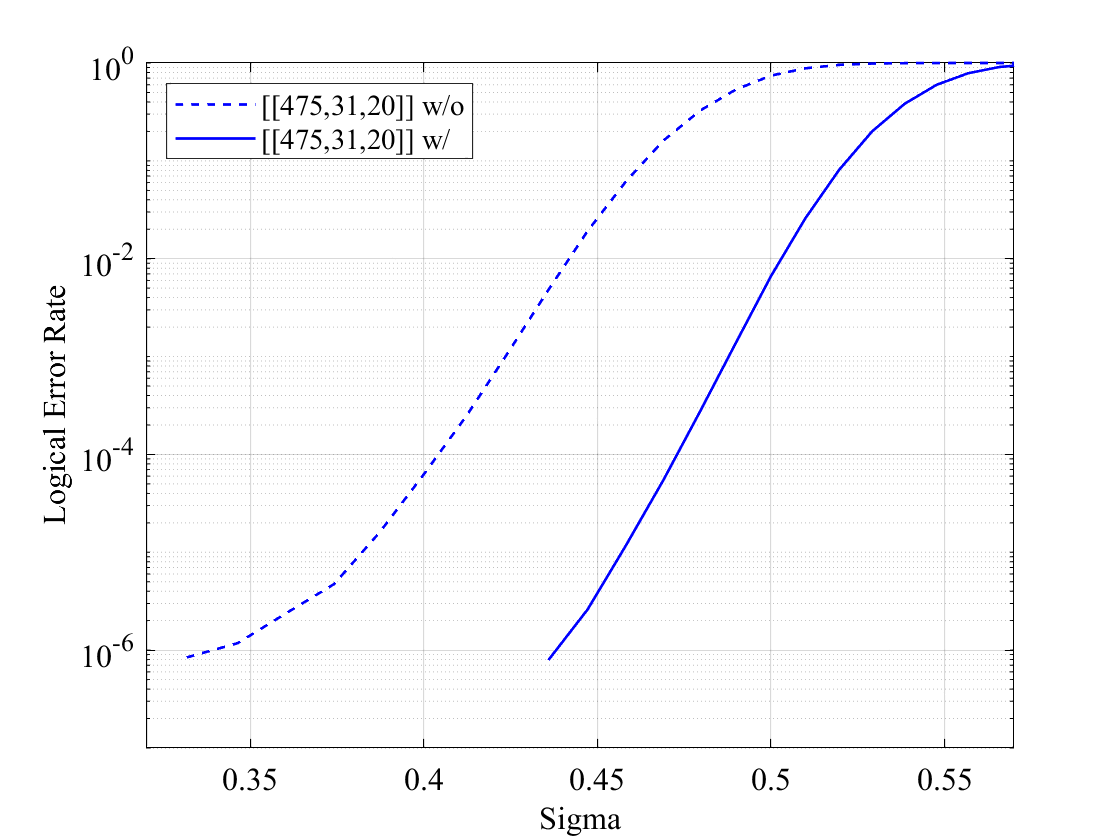}
    \caption{The LP code $\llbr 475,31,20 \rrbr$ from the LP04 family decoded with sequential min-sum without using the analog information exhibits the \emph{error flooring} effect. However, such performance degradation is not observed when we utilize the analog information from GKP error correction in the outer code decoder. The effect of analog information on error floors needs further investigation.
    }
    \label{fig:EF_plot_dmin20}
\end{figure}

The improvement in the decoding performance with the use of analog information is amplified at low logical error rates, specifically $10^{-4}$ and lower. This can be observed in Fig. \ref{fig:threshold_dv3dc4_with_withoutSoftInfoLayered} for all codes in the LP04 family. Classical LDPC codes are prone to a phenomenon called \emph{error floor} \cite{03Richardson,VNC_2014_Book} where the logical error rate curve flattens at low noise regime. This is attributed to small graphical configurations inside the Tanner graph referred to as \emph{trapping sets}. Recently, trapping sets of QLDPC codes have also been explored, and new code design and decoder improvement strategies avoiding small trapping sets have been discussed \cite{Raveendran-Quantum2021}. Here, using a particular code from the LP04 family, we demonstrate that the GKP analog information significantly improves the error floor problem. Intuitively, it appears that the ``analog'' knowledge about the error on each qubit seems to suffice to bias the decoder onto a particular pattern inside the trapping set. It is an intriguing phenomenon with no classical equivalent because classical error correction only employs channel information that is not qubit- and error-realization-specific. In Fig. \ref{fig:EF_plot_dmin20}, the error flooring effect is observed when the LP code $\llbr 475,31,20 \rrbr$ from the LP04 family is decoded with sequential min-sum without analog information. However, such performance degradation is not seen when analog information is used for the same code and decoding algorithm. We emphasize that this is a preliminary observation that requires further investigation.

\section{Conclusion and Future Directions}
\label{sec:Conclusion}

In this work, we demonstrated the concatenation of the GKP code with \emph{generic} QLDPC code families that are decoded using iterative decoding algorithms.
Specifically, we extended the recent investigations on surface-GKP schemes to the regime of \emph{finite rate} outer QLDPC codes, and proposed an explicit method to feed the GKP analog information into the outer code's iterative decoder.
The iterative decoder is the normalized min-sum algorithm (MSA) with a sequential node update schedule, and the MSA is widely deployed in classical error correction applications due to its hardware-friendly nature.
Since our focus is on exploring the gains of such a concatenation scheme, we considered a simple noise model where all GKP data qubits undergo a Gaussian random displacement and all GKP ancillas are noiseless (i.e., infinitely squeezed).
We chose two QLDPC code families arising from the lifted product construction for our simulations.
We showed that the MSA decoder is able to exploit the analog information from the inner GKP code error correction and significantly improve the noise threshold.
Furthermore, we showed that the GKP analog information combined with the sequential schedule enables the coding scheme to surpass the CSS Hamming bound. 
The contribution of the degeneracy of the QLDPC codes to this observation remains to be understood.
This concatenation scheme can be extended, opening the door to many possible outer QLDPC codes in this concatenated-GKP framework. 
Moreover, there is room for improvement in the iterative decoders used, especially if decoder design takes advantage of the degeneracy of QLDPC codes.

We note that the considered CSS-QLDPC-GKP architecture is based on the square-lattice GKP code. Our proposed scheme could be easily generalised to other rectangular GKP lattices, which could prove useful against noisy channels with biased noise in the $Q-P$ quadratures. Additionally, even for the symmetric Gaussian random displacement channel considered here, it might be beneficial to consider rectangular GKP lattices as well. For such an architecture the logical GKP errors will be biased, with the probabilities of the logical $X$ and $Z$ errors being now different. However, there exist QLDPC codes that are tailored to overcoming biased errors~\cite{roffe2022bias}. These codes could then be used together with rectangular-lattice GKP code in a similar spirit to the way the rectangular-lattice GKP code can improve performance against symmetric noise when concatenated with the surface code~\cite{Hanggli-pra20}. Since the lifted-product QLDPC codes tailored to biased noise are non-CSS~\cite{roffe2022bias}, further work will be required to adapt our decoding to this scenario. Another useful GKP lattice is the hexagonal one, which allows for the densest packing of circles~\cite{Gottesman-pra01} leading to better upper bounds on the logical error probability against the Gaussian random displacement channel than for the square-lattice GKP code~\cite{albert2018performance}. Moreover, the numerical optimisation in~\cite{noh2018quantum} also suggests that the hexagonal-lattice GKP code might be the optimal single-mode code against the pure loss channel. Since this non-rectangular lattice cannot be decomposed into a direct sum of two disjoint lattices along the $Q$ and $P$ quadratures respectively, this GKP code does not have the CSS property~\cite{Gottesman-pra01} and hence more work will be needed in order to develop decoding procedures for the corresponding QLDPC-GKP concatenated codes.   

In our simulation results, we observed that utilizing the analog information helps the iterative decoder to escape from the harmful trapping set configurations present in QLDPC codes, leading to no error floor or a significantly lower error floor. 
In our future works, we will analyze the effect of analog information on the trapping sets of QLDPC codes, which can result in better QLDPC codes and decoders.
We will also investigate if the additional analog information from inner codes can be used to exploit the degeneracy property of QLDPC codes in their iterative decoding \cite{Fuentes_DegeneracyImpact_IEEEAccess21}. 

Besides these specific results, our observations could trigger new research in decoder construction and analysis, e.g., on optimal thresholds under analog information, as well as in information-theoretic questions, e.g., capacity of (discrete variable) codes under analog information.
Even though it might not be possible to identify a reasonable classical setting where each received bit is equipped with its own analog information, it might be useful to start investigating the aforementioned questions under such a purely theoretical classical setting.
The experience gathered from such a setting could provide insights and benchmarks for the practically motivated quantum problem considered here.

Since we have established the utility of GKP analog information for iterative decoding of outer QLDPC codes under a simple noise model, a natural extension would be to consider more realistic noise models such as in the recent surface-GKP investigations~\cite{Noh-pra20,Noh-arxiv21}.
In particular, we will consider finitely squeezed GKP ancillas for syndrome extraction of both inner and outer codes.
In such a scenario, the analog syndrome itself would be noisy, which needs to be incorporated in the syndrome-based iterative decoder.
We expect the GKP-QLDPC concatenation scheme to work well even in such a noisy syndrome setting using outer code decoders that can utilize the soft syndrome information \cite{ImprovedDecodingSoftInfo_2021,softSynd_Decoder}.
Furthermore, having considered an application-agnostic setting in this paper, we will also consider noise models specialized towards fault-tolerant quantum computing or quantum communications (e.g., quantum repeaters~\cite{rozpkedek2021quantum}).

\section*{Acknowledgments}
We would like to thank David Declercq for help with QC-LDPC code construction and insights. This work is funded by the NSF under grant
NSF-ERC 1941583. The work of the University of Arizona team is also supported by the NSF under grants CIF-1855879, CIF-2106189,
CCF-2100013 and ECCS/CCSS-2027844.
The work of the University of Chicago team is also supported by the ARO (W911NF-18-1-0020, W911NF-18-1-0212), ARO MURI (W911NF-16-1-0349, W911NF-21-1-0325), AFOSR MURI (FA9550-19-1-0399, FA9550-21-1-0209), AFRL (FA8649-21-P-0781), DoE Q-NEXT, NSF (OMA-1936118, EEC-1941583, OMA-2137642), NTT Research, and the Packard Foundation (2020-71479).
Bane Vasi\'c has disclosed an outside interest in his startup company Codelucida to The University of Arizona. Conflicts of interest resulting from this interest are being managed by The University of Arizona in accordance with its policies.


\appendix
\section{QC-QLDPC Codes}
\label{sec:AppendixA}
The lifted product codes of LP04 family are obtained from the (3,4) regular QC-LDPC codes. For constructing the codes given in Table \ref{table:CodeParams34} with minimum distances - 10, 12, 18, and 20 respectively, we use Eq.~\ref{eq:LP_QLDPC_eq} with the base matrices as given below. 
Minimum distances for these QLDPC codes are computed by modifying the version of the error-impulse method \cite{impulseMethodClassical_Declercq}  for classical LDPC codes - adapted to suit for QLDPC codes.
In the following, the base protograph matrices are denoted as $B^L_{d_\text{min}}$ 
where the subscript $L$ denotes the circulant size.

\begin{align}
B^{7}_{10} & = 
\begin{bmatrix}
 0 &  0 &  0 & 0 \\
 0 &  1 &  2 & 5 \\
 0 &  6 &  3 & 1
\end{bmatrix}.
\label{eq_QC_dv3dc4_d10_L7}
\end{align}

\begin{align}
B^{9}_{12} & = 
\begin{bmatrix}
 0 &  0 &  0 &  0 \\
 0 &  1 &  6 &  7 \\
 0 &  4 &  5 &  2 
 \end{bmatrix}.
\label{eq_QC_dv3dc4_d12_L9}
\end{align}

\begin{align}
B^{17}_{18} & = 
\begin{bmatrix}
 0 &  0  &  0  &  0 \\
 0 &  1  &  2 &  11 \\
 0 &  8 &  12 &  13 
\end{bmatrix}.
\label{eq_QC_dv3dc4_d18_L17}
\end{align}

\begin{align}
B^{19}_{20} & = 
\begin{bmatrix}
 0 &  0  &  0  &  0 \\
 0 &  2  &  6 &  9 \\
 0 &  16 &  7 &  11 
\end{bmatrix}.
\label{eq_QC_dv3dc4_d20_L19}
\end{align}

Similarly, the lifted product codes of LP118 family are obtained from the $(3,5)$ regular QC-LDPC codes. For constructing the codes given in Table \ref{table:CodeParams35} with minimum distances - 12, 16, and 20 respectively, we use Eq.~\ref{eq:LP_QLDPC_eq} with the respective base matrices $B^{16}_{12}$, $B^{21}_{16}$, and $B^{30}_{20}$ as given below.

\begin{align}
B^{16}_{12} & = 
\begin{bmatrix}
 0 &  0 &  0 &  0 & 0 \\
 0 &  2 &  4 &  7 & 11 \\
 0 &  3 &  10 &  14 & 15
\end{bmatrix}.
\label{eq_QC_dv3dc5_d12_L16}
\end{align}

\begin{align}
B^{21}_{16} & = 
\begin{bmatrix}
 0 &  0 &  0 &  0 & 0 \\
 0 &  4 &  5 &  7 & 17 \\
 0 &  14 &  18 &  12 & 11
\end{bmatrix}.
\label{eq_QC_dv3dc5_d16_L21}
\end{align}

\begin{align}
B^{30}_{20} & = 
\begin{bmatrix}
 0 &  0  &  0  &  0 & 0 \\
 0 &  2  &  14 &  24 & 25 \\
 0 &  16 &  11 &  14 & 13
\end{bmatrix}.
\label{eq_QC_dv3dc5_d20_L30}
\end{align}
\end{document}